\documentclass[10pt]{nws}
\usepackage{afterpage,xcolor,charter,bm,graphicx,natbib,amsmath,amssymb,amsfonts,caption}
\usepackage[pdftex,pagebackref=true]{hyperref}
\usepackage{xy}\xyoption{all} \xyoption{poly} \xyoption{knot}  

\def\cD{\mathcal{D}}\def\F{\bm{F}}\def\G{\bm{G}}\def\W{\bm{W}}\def\bom{\bm{\omega}}\def\bth{\bm{\theta}}
					 \def\m{\bm{m}}\def\C{\bm{C}}\def\a{\bm{a}}\def\R{\bm{R}}\def\A{\bm{A}}
\def\eqno#1{eqn.~(\ref{eq:#1})}
\def\beginmat{ \left( \begin{array} }
\def\endmat{ \end{array} \right) }
\def\figsize{4.10in}
 
\def\figllgmdir{./}  
 
\title[Network Science]{Bayesian Dynamic Modeling and Monitoring \\ of Network Flows}  

\author[Chen et al.]
        {Xi Chen\footnotetext{\hrule\medskip{\em Corresponding author:} Xi Chen\href{mailto:chenxi199008@gmail.com}{\it  $<$chenxi199008@gmail.com$>$}, 
         LinkedIn Corporation,  Sunnyvale, CA 94085, USA. The research reported here was developed while Xi Chen was a PhD student in the Department of Statistical Science at Duke University.
         David Banks\href{mailto:David.Banks@duke.edu}{\it  $<$David.Banks@duke.edu$>$} is Professor, and Mike West\href{mailto:Mike.West@duke.edu}{\it  $<$Mike.West@duke.edu$>$} is The Arts \& Sciences Professor of Statistics \& Decision Sciences in the Department of Statistical Science at Duke University, Durham, NC 27708.
         \\ {\em Acknowledgements:} The authors are grateful for discussions with Mark Lowe and Jewel Thomas on aspects of the research reported here.  Maxpoint Interactive Inc. (now Valassis Digital) provided the network flow data of the reported case study as well as partial support 
         for the research. We also acknowledge the constructive comments of the editor, associate editor and three anonymous referees on the original manuscript. }\\
         LinkedIn Corporation,   Sunnyvale, CA 94085, USA\\ 
         \email{chenxi199008@gmail.com, David.Banks@duke.edu, Mike.West@duke.edu}}
 \author[Chen et al.]                 
        {David Banks \& Mike West\\
        Department of Statistical Science, Duke University, Durham, NC 27708-0251, USA\\ \phantom{.} }

\jdate{\today}\pubyear{2019}\pagerange{\pageref{firstpage}--\pageref{lastpage}}\doi{}
 
\begin{document}

\label{firstpage}

\maketitle 
\begin{abstract}
In the context of a  motivating study of dynamic network flow data on a large-scale e-commerce web site, we develop Bayesian models for on-line/sequential analysis for monitoring and adapting to changes reflected in node-node traffic. For large-scale networks, we customize core  Bayesian time series analysis methods using dynamic generalized linear models (DGLMs).
These are integrated into the context of multivariate networks using the concept of decouple/recouple that was recently introduced in multivariate time series. This method enables flexible dynamic modeling of flows on large-scale networks and exploitation of partial parallelization of analysis while maintaining coherence with an over-arching multivariate dynamic flow model. 
This approach is anchored in a case-study on internet data, with flows of visitors to a commercial news web site defining a long time series of node-node counts on over 56{,}000 node pairs. 
Central questions include characterizing inherent stochasticity in traffic patterns, understanding node-node interactions, adapting to dynamic changes in flows and allowing for sensitive monitoring to flag anomalies. 
The methodology of dynamic network DGLMs applies to many dynamic network flow studies. 
\\ \phantom{.} 
\end{abstract}

\newpage

\section{Introduction}
 
Key areas of network science have successfully adopted and refined statistical modeling ideas 
and methods from other fields, and stimulated new statistical developments of broader use.    The general area of dynamic networks has generated broad interest in questions of modeling changes in time of network (and graph) structures from statistics and machine learning perspectives; we comment on a number of recent contributions~\citep[including][]{hanneke2010discrete,richard2014link,newman2004analysis,newman2018network,holme2013temporal,holme2015modern,giraitis2016estimating,bianchi2018modeling}, and  relevant  Bayesian approaches~\citep[including][]{kim2018review,sarkar2007latent,xing2010state,xu2014dynamic} in Section~2.1.    

Our interest is in studies of time-varying patterns of integer counts of traffic flowing between nodes in a given network (as well as into and out of the network). 
This topic has broad application but has not previously been well-addressed in statistical terms. Past work on so-called network tomography~\citep[e.g.][]{Tebaldi1998} and physical
traffic flow forecasting~\citep[e.g.][]{Tebaldi2002,Queen2009, AnacletoEtAl2013a,
  AnacletoEtAl2013b} are relevant, but our dynamic network flows contexts present challenges that require new modeling approaches.  We exploit perspectives of Bayesian time series modeling and analysis to advance practicable methodology for characterizing and monitoring dynamic network flows. The motivating context is in internet traffic in e-commerce web-sites, where we 
present aspects of a case study based on the new modeling and analysis framework. 

Increasing access to streaming traffic data on networks drives interest in models to quantify inherent levels of variation in flows of traffic between network nodes and into/out of the network from other sources.  There are two aspects of this. The first is natural and unpredictable variation, the second is that of stochastic but sustained patterns in underlying trends over time.  State-space models directly address this: they couple 
observation (noise) models for unpredictable variation with latent state processes representing the  structure and relationships among nodes---and   patterns of time-variation in these relationships. 
 Understanding and appropriately characterizing normal patterns of variation in traffic flows is necessary to coherently address questions of monitoring   flows for potential anomalies, and then to intervene or otherwise respond to interesting inferred changes.

Key technical challenges are to develop real-time/sequential analysis that is computationally efficient 
and scalable with network size and data sampling rates. We address these questions from a new perspective of Bayesian time series analysis, 
introducing to the network science literature approaches based on 
dynamic generalized linear 
models (DGLMs) of proven utility in other fields for many years~\citep[e.g.][Chapter 14]{migon1985application,west1985dynamic,WestHarrison1997}. 
The class of DGLMs integrates non-Gaussian sampling models of traditional generalized linear models with state-space evolution models for time series. 
The subclass of models based on conditional Poisson sampling distributions is especially relevant to dynamic network flow studies as it its to other problems involving monitoring and forecasting multivariate systems of time series of counts~\citep[e.g.][]{BerryWest2018DCMM,BerryWest2018TSM}.

We extend basic univariate DGLMs to the multivariate dynamic system generated by large-scale networks. We do this using the modeling concept of {\em decouple/recouple} originally introduced for multivariate time series in financial and economic applications~\citep{GruberWest2016BA,GruberWest2017ECOSTA}, and that has been recently extended for network data with  simple network flow models~\citep{xi2017bdfm}. 
The latter reference represents our starting point here. The decouple/recouple idea is combined with the use of Bayesian model emulation, which maps inferences from collated sets of individual node-node flow models to an integrated multivariate network. This model enables us to explore dynamics in network-level activity intensity, node-specific flow dynamics, and node-node interactions over time with information from individual flow levels.

This new model class is quite general in admitting a range of possible state-space models for link-specific flow evolutions. In our case study, we utilize one of the most important special cases of (local) linear growth models for adapting to unpredictable changes in trends underlying flow patterns. Other applications will involve special cases customized to context-- such as dynamic regressions on known predictors, seasonal structures, and other Bayesian state-space structures used in time series analysis~\citep[e.g.][]{WestHarrison1997,PradoWest2010}. 
 
Our case study concerns internet traffic in e-commerce, where the flow data are counts of visitors moving between nodes that are clusters of web pages corresponding to meaningful categories in a commercial web site. The network example has over 56{,}000 node pairs and data for 288 time points within one day. Online advertisers are interested in many statistical issues related to traffic flow and 
site-segment content. The field has become quite sophisticated, and many methods have been 
employed to explore related problems, for instance, complex recommender 
systems~\citep{Koren:2009}, sentiment analysis~\citep{Pang:2008}, and text mining 
\citep{Soriano:2013}, etc. However, basic questions of statistical modeling to 
characterize, monitor and potentially understand dynamics in traffic across site-segments have not received 
as much attention as they warrant. In particular, there is potential commercial value as well as inherently interesting methodological concern in identifying and adapting to fluctuations in the changes of a 
site's popularity on short time scales. There is related interest in understanding the interactions between sites with respect to traffic, and in model-based monitoring for subtle anomaly detection in sub-networks. In such applications, one important focus is \lq\lq ads pacing'' which relates to the speed with which advertisers deplete the budget assigned to a specific ad. A good understanding of traffic dynamics can help control the delivery of ads and budget spending, and thus improve the matching between demand-side and supply-side to maximize revenue.

We extend the recent work of~\cite{xi2017bdfm} in terms of network context and goals. 
We introduce the rich class of DGLMs for latent Poisson rates, with opportunities to substantially advance methodology for time-varying flow characteristics  on larger networks.  In addition to extending the statistical modeling methodology, analysis of historical patterns in flow rates is enabled with extensions of existing Bayesian time series analysis with DGLMs to include retrospective posterior sampling of state vectors over time.  This underlies the ability to map inferences on the time-varying parameters of so-called dynamic gravity models to inference on global network flow rates, dynamics in node-specific rates, and dynamics in node-node relationships.  The application in this paper involves a large network that highlights the utility of the new models, along with the technical advances in 
modeling and computation for larger problems.

Section~\ref{sec:notation} describes the network context, basic statistical setting and notation. It outlines the concept of decoupling multivariate flows into those on univariate node-node pairs, as well as dynamic parameter mappings for inference on node-specific and node-node interactions. 
Section~\ref{sec:dglm} details the class of conditionally Poisson DGLMs in a general setting, and then focuses on the specific example of 
local linear growth models for latent flow rates underlying network counts.  Linked to these models, the Appendix 
has three components. Appendix~A summarizes technical details for mapping of DGLM inferences to the context of dynamic gravity models. Appendix~B summarizes the standard on-line learning algorithm for DGLMs and a novel methodological extension required for approximate posterior simulation of full time trajectories of latent state vectors. Appendix~C provides further examples of inferred dynamics in node-node interaction 
 effects, illustrating  potential insights for online advertising.

Our case study involves data from the Fox News web site, where flows represent individuals browsing web pages. Section~\ref{sec:data} introduces the 
data and context,   develops initial DGLM analysis and discusses some comparisons with the prior approach in~\cite{xi2017bdfm}.
Section~\ref{sec:dgm} discusses the mapping of posterior distributions from the DGLM analysis results to the time-varying parameters of 
highly structured dynamic gravity models (DGMs), the latter then defining inferences on node-specific and node-node interaction effects and how they vary over time. Several highlights are discussed using specific nodes from the news web site.   We give some concluding comments in 
Section~\ref{sec:closing}.

\section{Background}
\subsection{Related Work}

Dynamic network analysis is an increasingly active research area and includes a multitude of network contexts and analysis goals.   In statistics, for example, \cite{hanneke2010discrete} proposed a family of statistical models extending traditional exponential random graph models (ERGMs) to a dynamic setting, with interest in potential problems of testing for changes as well as classification.  Statistical and machine learning approaches have evolved to address the problem of link prediction in time-evolving graphs~\citep[e.g.][]{richard2014link}.  Perspectives from different areas of the physical sciences have included approaches based on  graph theory~\citep[e.g.][]{newman2004analysis,newman2018network,holme2013temporal,holme2015modern}, while dynamic network models are increasingly visible in applied studies in areas including biology~\citep[e.g.][]{uetz2000comprehensive,giot2003protein}, econometrics and marketing~\citep[e.g.][]{giraitis2016estimating,bianchi2018modeling}.

Various Bayesian statistical models have also been developed for dynamic network studies with state-space models~\citep{kim2018review}. State-space models represent data with coupled observation (noise, natural random variation) components and latent process (underlying state variables with stochastic but sustained patterns over time). Conditionally linear, Gaussian models are amenable to analytic computation using Kalman filtering-style analysis, and more extensive analysis using broader classes of Bayesian state-state space models~\citep[e.g.][]{WestHarrison1997}. However, many/most problems of network inference in dynamic contexts involve non-Gaussian data structures and non-linear effects, so one needs stronger computational methods such as Markov chain Monte Carlo~\citep{xu2014dynamic,hoff2011hierarchical}.  
For example, \cite{sarkar2007latent} map the dynamic co-occurrence data to dynamic embeddings in low-dimension Euclidean space, while~\cite{xing2010state} and~\cite{xu2014dynamic} extend well-known stochastic block models to study dynamic network tomography.  More closely allied to some of our main interests here are past works on the so-called 
network tomography problem~\citep[e.g.][]{Tebaldi1998} and state-space models for physical
traffic flow forecasting~\citep[e.g.][]{Tebaldi2002,Queen2009, AnacletoEtAl2013a,
  AnacletoEtAl2013b}.  Our work presents a broader approach to such studies by exploiting rich classes of Bayesian state-space models to characterize and analyse complex, time-varying flows on networks.

This paper extends the recent work of~\cite{xi2017bdfm} in terms of network context and goals.  Using an example of a small network (20 nodes), that work utilized a relatively simple smoothing model for latent time-varying Poisson rates on node-node pairs, and introduced the key idea of decouple/recouple to enable scaling. While based on a state-space approach, the simple smoothing model  is simply not able to adequately capture trends or abrupt changes in network flows, and the model structure is not flexible enough to incorporate other node/flow features. Those challenges are addressed by DGLMs whose utility has been proven in other fields for many years~\citep[e.g.][Chapter 14]{migon1985application,west1985dynamic,WestHarrison1997}.

\subsection{Network Structure and Notation} 
\begin{figure}[htbp!]
$$
\xymatrix{ 
\textrm{In-flow} \ar@/_0pc/_*+{x_{0it}}[rrrd]  & & & & & & *+++[o][F-]{1} \ar@{.}[d] & \\
& & &
	*+++[o][F-]{i} \ar@/^/@{-->}^*+{x_{i1t}}[rrru] 
	\ar@{-->}^*+{x_{ijt}}[rrr] \ar@/_/@{-->}^*+{x_{iIt}}[rrrd] 
	\ar@/_0pc/@{-->}_*+{x_{i0t}}[llld] & & & *+++[o][F-]{j} \ar@{.}[d] & \\
 \textrm{Out-flow} & & & & & & *+++[o][F-]{I} & \\
}
$$ 
\caption{Network schematic and notation for flows at time $t$.}
\label{fig:networkcartoon}
\end{figure}
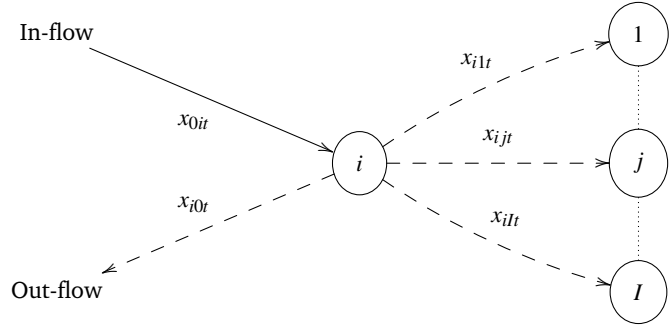

Consider a closed network defined on $I$ nodes with counts of traffic flowing between pairs of nodes observed sequentially over equally-spaced time 
$t=1,2,\ldots$.
This defines a highly interdependent, multivariate count time series. The counts represent  units who individually enter the network at one of the nodes at some specific time, transit to other nodes, may stay at a node for a period of time, and may exit the network at some time point. The case study involves IP addresses (indicating individuals) browsing a web site comprised disjoint of sets of web categories
(e.g., Arts \& Entertainment, Weather) so that these sets are the nodes. In this and other applications, 
an additional node indexed by 0 is needed to represent flows into, or out from, nodes in the core network. 
At each time point $t=1{:}T$, let $n_{it}$ be the number of occupants of 
node $i$, and let $x_{ijt}$ be the flow count from node $i$ to $j$, 
including the in-flows $x_{0it}$ and out-flows $x_{i0t}$, as shown in 
Fig.~\ref{fig:networkcartoon}. 

\section{Statistical Structure \label{sec:notation}}
\subsection{Dynamic Poisson and Multinomial Flow Distributions} 
The natural class of dynamic models have hidden Markovian structures with latent rates defining transitions between nodes over time, 
the rates themselves being time-varying.  We adopt conditionally independent Poisson models ($Poi$)
for in-flows to the network coupled with  conditionally independent multinomials ($Mn$)
for flows from each node, all with time-varying parameters.  That is, 
for $t = 1{:}T$, and network nodes $i = 1{:}I$,
\begin{equation}
x_{0it}|\phi_{0it} \sim Poi(\phi_{0it}) \quad\mbox{and}\quad
\bm{x}_{i\cdot t}|n_{i,t-1},\bm{\Theta}_{t} \sim Mn(n_{i,t-1},\bth_{it}). 
\label{eq:poissmultin}
\end{equation}
where $x_{i\cdot t}$ is the vector of outflows to all $I$ nodes from node $i$ at time $t.$  
In the notation here, $Poi(\phi_{0it})$ denotes a Poisson distribution with mean $\phi_{0it}$, 
and $Mn(n_{i,t-1},\bth_{it})$ denotes a multinomial distribution with $n_{i,t-1}$ trials and $I+1$ cell probabilities in the vector $\bth_{it}$.
The defining parameters of these distributions are time $t$ values of underlying latent processes: 
$\phi_{0jt}$ is the time-varying rate process governing flows  into node $j$ from the external node 0; 
$\bth_{it}=(\theta_{i0t}, \theta_{i1t}, \cdots, \theta_{iIt})'$ is the vector of
time-varying transition probabilities from node $i$ to all other nodes 
with $j = 0$ indicating departure from the network.  

Modeling flexibility and computational efficiency are key needs in large scale dynamic network analysis.
In our case study, one goal of the end-user is to optimize online advertisement placement 
by analyzing browsing patterns, and these decisions must be made in less than a  second, demanding fast and effective analysis.
The theory underlying multinomial models allows us to address this using decoupling of flow transitions to node-node pairs, enabling use of univariate 
DGLMs in parallel, followed by theoretical recoupling for exact inference on the sets of multinomial probabilities. 
In the decoupling step, network flows originating from nodes $i=1{:}I$ within the network are implicitly conditionally 
independent Poisson random variables, denoted by 
\begin{equation}
x_{ijt}|\phi_{ijt} \sim Poi(m_{it}\phi_{ijt}) \quad\mbox{where}\quad m_{it} = n_{i,t-1}/n_{i,t-2},
\label{eq:occratio}
\end{equation}
where: {\em (i)}  the $\phi_{ijt}$ are latent Poisson rates governing the outflows from node $i$ to all other nodes $j=0{:}I$ at time $t,$ and 
{\em (ii)} the {\em occupancy ratio} $m_{it}$ 
provides appropriate scaling of outflows from node $i$ as its level of occupancy counts changes over time.

Decoupling allows individual flows to be updated independently, 
to achieve fast parallel computing per unit of observation time. 
One aspect of recoupling is to then directly  revert to the fundamental time-varying multinomial transition probabilities of \eqno{poissmultin} via 
$
\theta_{ijt} \propto m_{it}\phi_{ijt} \propto  \phi_{ijt} $ subject to summing to 1 over $j=0{:}I.$ Given any set of the $\phi_{ijt}$ simulated from posterior distributions, this trivial computation provides inferences on the multinomial transition probability processes.  

\subsection{Mapping to Dynamic Gravity Model \label{sec:mapptoDGM} } 

As for decoupling, besides mapping to multinomial transitions as discussed in Section 2.3, the second aspect is to map inferences on the $\phi_{ijt}$ to those of a separate {\em dynamic gravity model} (DGM~\cite[e.g.][]{West1994, Sen:1995,Cogdon2000}). This mapping enables evaluation of flow patterns at various levels of the network: {\em (i)} overall network level, {\em (ii)} the level of each individual node, and {\em (iii)} for all pairs of nodes. The details follow~\cite{xi2017bdfm} are summarized here, with some additional technical detail in Appendix~A. 
A dynamic gravity model (DGM) represents $\phi_{ijt}$ as a product of four terms, at each $t$ and for each pair of nodes, viz
\begin{equation}\label{eq:DGM} 
 \phi _{ijt} = \mu _t  \alpha _{it}  \beta _{jt}  \gamma _{ijt}
\end{equation}
for each within-network node $i=1{:}I$ and all  $j=0{:}I$ over all $t=1{:}T.$ 
Here the latent rates $\phi_{ijt}$ are mapped to: {\em (i)} a network-level flow rate process $\mu_t$;  {\em (ii)} a multiplicative {\em origin effect process}  
$\alpha_{it}$  for each node $i;$  {\em (iii)}  a multiplicative {\em destination effect process} $\beta_{jt}$ for each node $j;$ and 
{\em (iv)} an {\em affinity process} $\gamma_{ijt}$, a dynamic interaction term representing the directional attractiveness of node $j$ as a destination for flows from node (i) relative to the contributions of baseline and main effects. 

Given posterior samples  of the $\phi_{ijt}$ over nodes and time, we can directly map to the DGM parameters for more incisive inference on node-specific and node-node interactions over time.  In this sense, the flexible class of decoupled/recoupled DGLMs can be used as a Bayesian emulator for inference in the DGM. 
Technical details of the mapping are given in Appendix~A.

\section{Dynamic Generalized Linear Models (DGLMs)  \label{sec:dglm}}

DGLMs are generalized linear models~\citep{mccullough1989generalized,west1985generalized}  with time-varying parameters defined by state-space evolutions of regression vectors. 
Time series observations are conditionally drawn from a sampling model in the exponential family, and the natural parameter of the distribution is regressed on a time-evolving state-vector. These models build on dynamic linear models and are central in Bayesian work in applied time series with non-Gaussian data~\citep{WestHarrison1997,PradoWest2010}.   We focus here on the special example of conditionally Poisson models for both the network inflow count time series 
and the decoupled within-network flow series, i.e.,   $x_{ijt}$ for all node pairs $i,j\in 0{:}I$ and all times $t=1{:}T$ for which the sampling models are
$x_{ijt}|\phi_{ijt} \sim Poi(m_{it}\phi_{ijt}) $ with $m_{0jt}=1$ for inflows from outside the network.   DGLMs are created via state-space models for the latent rate processes $\phi_{ijt}$.

Ignoring the node indices $i,j$ for clarity, consider a single Poisson rate process $\lambda_t$. Suppose that 
$\lambda_t = \log(\phi_t)$ is defined via an underlying linear, state-space, Markov model in which 
\begin{equation} 
\lambda_t = \F'_t\bth_t \quad\mbox{and}\quad 
\bth_t = \G_t\bth_{t-1}+\bom_t \quad\mbox{where}\quad  \bom_t\sim [0,\W_t],
\label{eq:DGLMevolution}
\end{equation}
with the following elements:   {\em (i)}  $\F_t$ is a known $p\times 1$ regression vector of constants and/or known values of predictors at time $t$; 
{\em (ii)} $\bth_t$ is the corresponding $p\times 1$ dynamic regression parameter vector, known as the state-vector, at $t$; 
{\em (iii)} $\G_t$ is a known $p\times p$ state-evolution or transition matrix; {\em (iv)}  $\bom_t$ is a random $p\times 1$ innovation vector representing stochastic changes to the state at time $t$; {\em (v)} the $\bom_t$ are independent over time, and the notation indicates $\bom_t$ is zero mean and has known variance matrix $\W_t.$ 
 
Model specification depends on context, of course, and there are  widely used subclasses in which $\F_t$ and $\G_t$ take specific forms\citep[e.g.][Chapter 4]{PradoWest2010}. Some examples include DGLMs when $\F_t$ includes values of known covariates (predictors), intervention indicators, and constants representing groups and design variables, in which cases the corresponding entries in $\bth_t$ are dynamic regression coefficients. Natural evolution models then have corresponding rows of $\G_t$ as zero but for the implied column index, so that the model indicates a random walk time evolution for those parameters.  Relevant to many applications are examples where both $\F_t\equiv \F$ and $\G_t\equiv \G$ are constant with specific forms chosen to define local smoothing and interpolation, such as in models $M_1$ and $M_2$ defined by 
\begin{equation*} 
\begin{aligned} 
M_1: & \quad  \F = \beginmat{c}1\\0\endmat\quad\mbox{and}\quad\G = \beginmat{cc} 1&1\\ 0&1 \endmat,\\
M_2: & \quad   \F = \beginmat{c}1\\0\\0 \endmat\quad\mbox{and}\quad\G = \beginmat{ccc} 1&1&0 \\ 0&1&1 \\ 0&0&1 \endmat. 
\end{aligned}
\end{equation*}
In model $M_1$, the  latent state vector $\bth_t = (\lambda_t, \rho_t)'$ 
consists of the current level  of the latent $\lambda_t$ process and the time $t$ change in level (the discrete gradient, or \lq\lq growth'') term $\rho_t$. This is a local linear growth model (LLGM) and one of the most widely used DGLMs both alone or as a component of more elaborate models. The model defines local linear interpolation of time-varying trends that are otherwise regarded as unpredictable, and is key to retrospective smoothing of patterns in time series. Model $M_2$ is a more elaborate local quadratic model in which the third element of the state vector represents time-varying changes in gradient. More complicated local smoothing can be defined by higher-order polynomial DGLMs with obvious extension~\citep[][Chapters 7,10]{WestHarrison1997}.   The case study of this paper adopts the class of LLGMs defined by $\F,\G$ as in model $M_1$ above. 

Summary details-- including algorithms for implementation-- of Bayesian analysis of general DGLMs is given in Appendix~B. This includes
details of sequential learning, i.e., forward filtering to process data as it arrives and sequentially update prior-to-posterior summary information for the state vectors $\bth_t$ over time.  At any time $t,$ this enables inference on the current state and forecasts of coming data.  This online analysis is most relevant to sequential learning and monitoring of flows in many applications.  Then, based on an observed time series of flows over a period of time $1{:}T$,  key interests are addressed by  retrospective analysis that examines inferences on historical trajectories of state vectors, and any functions of them of interest. Bayesian analysis here is best addressed using simulation of posteriors over historical trajectories, and the implied posteriors for past evolution in patterns of substantively interesting parameters such as those of dynamic gravity models than can be implied.    Full technical and algorithmic details of this are summarized in Appendix~B.  
One key element of model specification is the extent and nature of time-variation in the state vector as defined by the variance matrices $\W_t.$ 
These are specified using the standard discount factors method; see Chapter 6 of~\cite{WestHarrison1997} and Section 4.3.6 of~\cite{PradoWest2010}, and the additional technical details in Appendix~B.

\section{LLGM Analysis of Fox News Flow Data \label{sec:data} }
\subsection{Fox News Flow Data}

The case study concerns flow data recording individual visitors (in terms of IP addresses) to well-defined nodes of the Fox News website. 
Our sample of data here concerns traffic on  September 17th, 2015 (a Thursday) segmented to IP addresses linked to visitors from only the 
Eastern Daylight Savings time zone.  The website is structured  by the Adex Category, a partition derived from 
text mining the webpage content and widely used in online advertising for webpage  analysis. 
There are $2{,}208$ pre-defined categories, including $26$ main categories 
and different levels of sub-categories. 
The $26$ main categories are Arts \& Entertainment, Computers \& Electronics, 
Finance, Games, Home \& Garden, Business \& Industrial, Internet \& Telecom, 
People \& Society, News, Shopping, Law \& Government, Sports, 
Books \& Literature, Real Estate, Beauty \& Fitness, Health, 
Autos \& Vehicles, Hobbies \& Leisure, Pets \& Animals, Travel, 
Food \& Drink, Science, Online Communities, Reference, Jobs \& Education 
and World Localities. 
Although there can be web page content that combines different categories, 
the Adex Category tool enforces a strict partition.
Exploratory study found little cluster structure among subcategories sharing 
the same main category.

Data are aggregated to five-minute intervals, suggested by  stability of exploratory analysis results
across temporal levels of data aggregation.
This defines a time series with $T=288$ time points having the structure  
described in Section \ref{sec:notation}. 
At each time interval, over the directed network with nodes classified by 
Adex Category, the data include counts of transitions of visitors between each pair, 
incoming flows from outside the Fox News website to each node, 
as well as the total number of people visiting each node.  There are no relevant additional covariates available, so the analysis focuses wholly on 
temporal trends in network, node-specific and node-node interactions as evidenced through analysis of flexible DGLMs that allow and adapt to changes over time. 
This is done using the special case of local linear growth DGLMS, i.e., the LLGM framework.

Most people spend no more than five minutes on a single 
web page~\citep{Jansen2007}.
Therefore, visitors who spend more than five minutes in a node
are deemed inactive and handled as if they have left the website.
Also, user information is unavailable before and after 
September 17, so the inactivity rule means the first and last five 
minute intervals are eliminated from the time series, 
which now has length $T=284$.
Then, there are some 
categories with little or no traffic during the entire day, so only those 
categories with sufficient data are considered in the analysis. 
By applying a threshold of $1$ for the total traffic across all $T = 284$ 
time periods, $I=237$ out of an initial superset of $2{,}208$ categories are left for analysis.

\subsection{Some Initial DGLM Analysis Summaries}

\begin{figure}[htbp!]
    \centering 
    \includegraphics[width=\figsize]{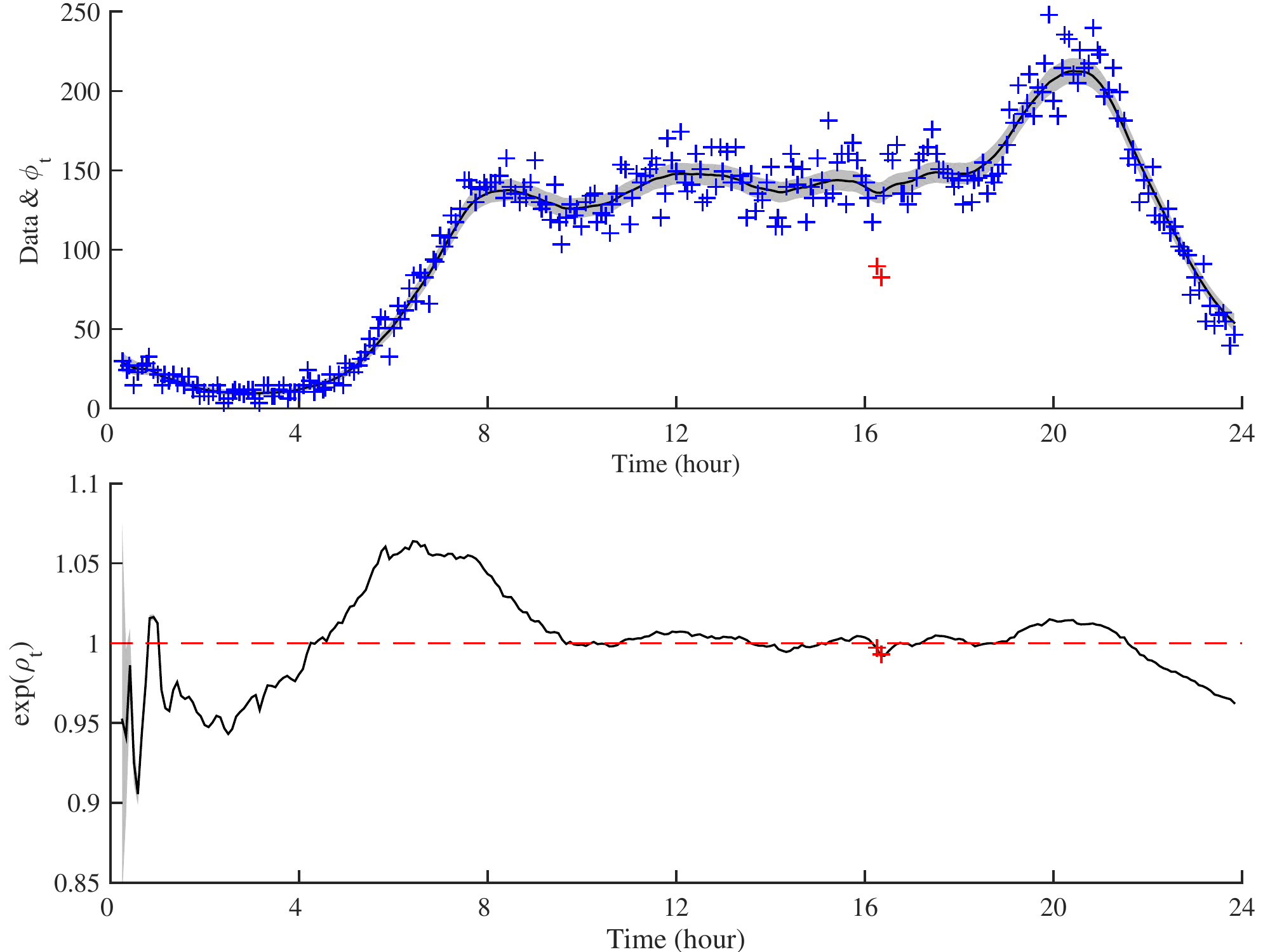} 
  \caption{LLGM  forward filtering of counts staying at node $i=$ Arts \& Entertainment on September 17, 2015. {\em Upper}: data (crosses),  filtered mean (solid) and $95\%$ CI (shaded) of trajectories of $\phi_{i,i,t}$; {\em Lower}: filtered mean (solid) and $95\%$ CI (shaded) of local linear growth term $\rho_{i,i,t}$. Two low-count  values denoted by  red '+'s are mentioned in the text. } 
 \label{fig:llgmff}   
\bigskip\bigskip
      \includegraphics[width=\figsize]{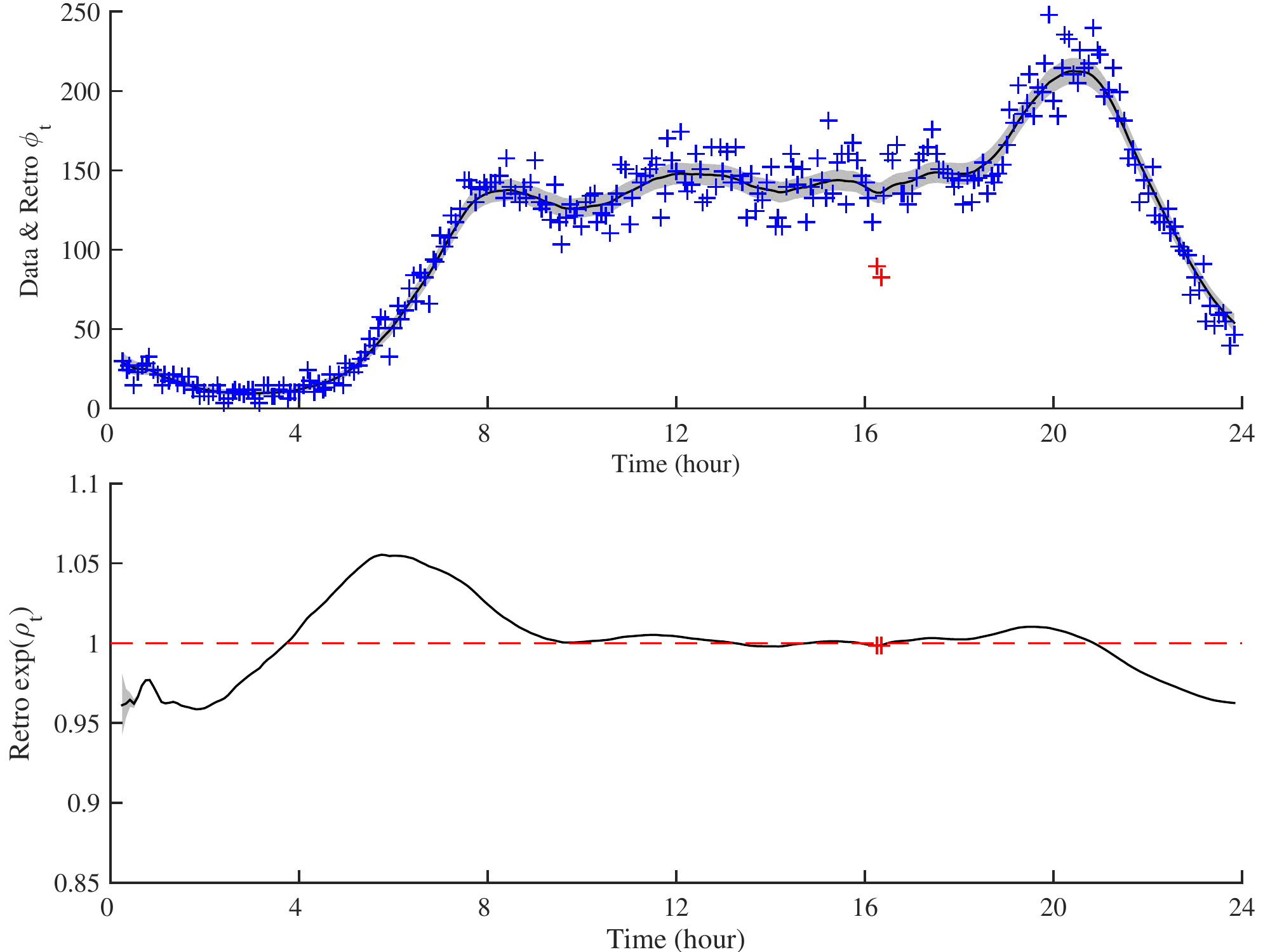} 
  \caption{As in Fig.~\ref{fig:llgmff}, but now showing full retrospective analysis. }
 \label{fig:llgmbs}        
\end{figure}

We focus on individual flows, and evaluate performance by comparing the 
accuracy of the one-step-ahead predictions against predictions from the 
Bayesian Dynamic Flow Model (BDFM) of ~\cite{xi2017bdfm}.
In our LLGM analysis, for each individual flow, the latent state vector 
$\bth_{ijt}$ has two components: the rate parameter on the 
log scale $\lambda_{ijt}$ and the local linear growth term $\rho_{ijt}$. 
The prior on $\lambda_{ij0}$ is chosen with mean equal to the point estimate based on a simple average of data 
in the five minutes prior to the beginning of the time series, and the 
prior mean for $\rho_{ij0}$ is $0$. 
The initial prior variance matrix has zero covariances and diagonal entries $0.1$. The levels of prior variance represent a relatively diffuse initial prior that allows for swift adaptation to the data over the initial few time points.
The discount factor for all reported analyses is set as $\delta=0.9$, for all node pairs. This level of discounting encourages smoothness 
but also allows the model to flexibly adapt to changes, and variations (that have been explored) might marginally improve local descriptions for some node pairs but are secondary 
to the interests and emphases here.

Decoupling allows us to apply LLGM simultaneously to all of the flows. 
The LLGM analysis of flows staying at the category Arts \& Entertainment
is used to illustrate the model performance. 
Fig.~\ref{fig:llgmff} shows the forward filtering results for both the 
rate parameter and the local linear growth term, while Fig.~\ref{fig:llgmbs} 
shows the results as smoothed by backward sampling. 
In general, both the sequential and retrospective analyses capture patterns of change over time  
well,  and the efficacy of retrospective analysis is highlighted.    
We note that the intervals representing uncertainty about trajectories are very tight, indicating a high level of precision in estimating the underlying transition processes.  This is true in several of the following graphs for other model parameters. 

There are periods when volatile patterns in the data challenge model adaptability. 
For example, in the morning ($8$:$00$ - $10$:$00$) and late at night 
($20$:$00$ - $24$:$00$), forward filtering analysis without any intervention underestimates the swings in the rate 
parameter $\phi_t$. 
The sequential analysis is more sensitive to outliers. 
For instance, at around $16$:$00$, there are two data points with 
low counts compared to the data before and after; see the red '+'s in Figs.~\ref{fig:llgmff} and~\ref{fig:llgmbs}. 
These two data points drive the rate down in the forward filtering. 
That said,  the retrospective analysis is able to resolve these two issues by 
using information from the later in the series.
Looking back, retrospective posterior analysis provides smoother and more accurate inference on the rate parameter.

Forward filtering and backward smoothing for the local linear growth term 
give insight into how trends vary during the day. 
For example, at first the Poisson rate for people staying in Arts \& Entertainment  
decreases at $0$:$00$ - $4$:$00$, but its decrease slows, reaching the minimum at around 
$4$:$00$ a.m. 
Afterwards, the rate increases rapidly and reaches the first peak of the day at around 
$8$:$00$ a.m., then maintains a steady, high level until around $19$:$00$. 
The $\phi_{ijt}$ then reaches its second peak at around $20$:$00$, and after that, the 
activity level declines rapidly until midnight.

\subsection{Comparison with BDFM}

\begin{figure}[htbp!]
    \centering 
    \includegraphics[width=\figsize]{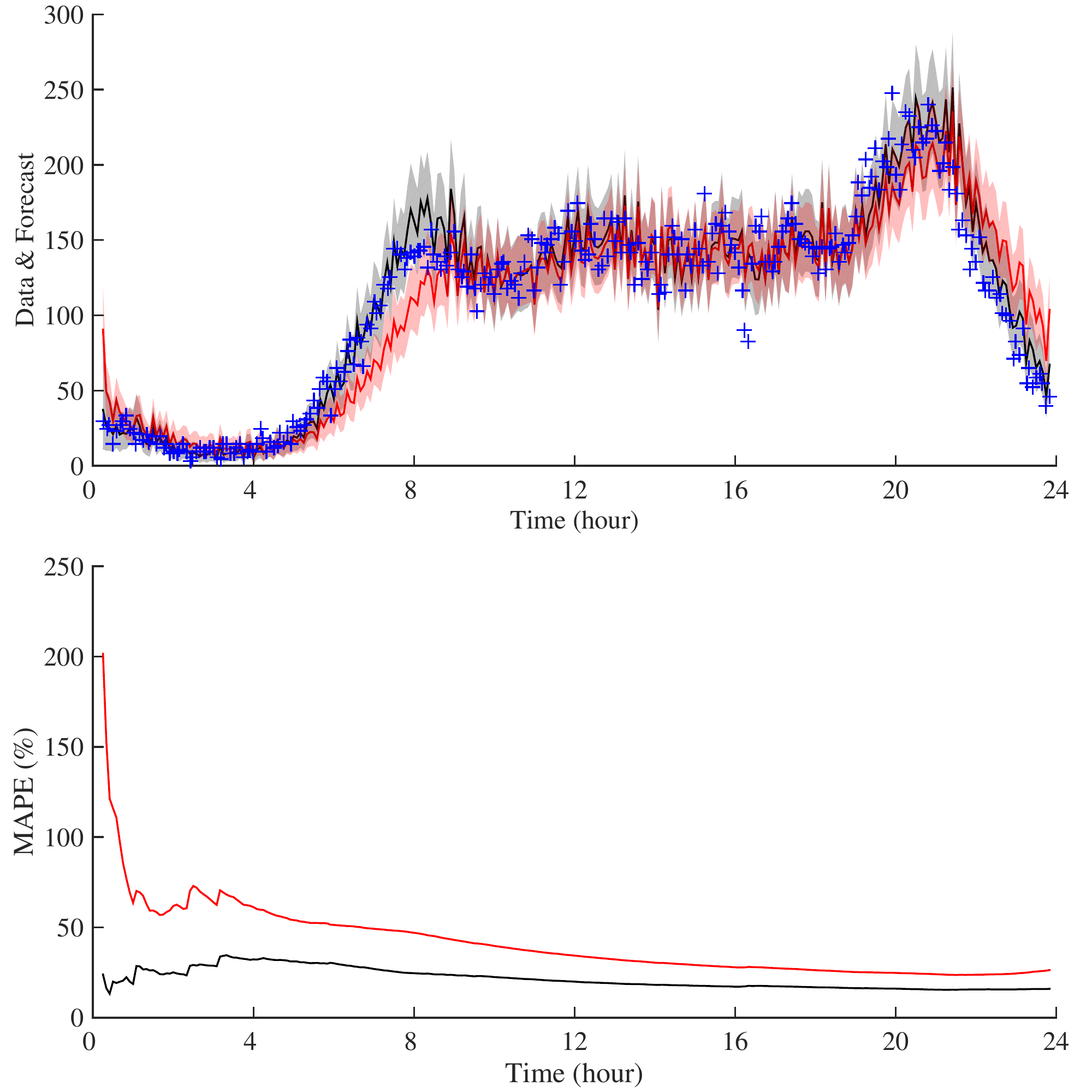} 
  \caption{Summary of analyses of counts staying at node $i=$ Arts \& Entertainment using both LLGM and BDFM. 
Data are from September 17, 2015.  
{\em Upper}: Data (+) with one-step forecast means and 95\% intervals from the 
	LLGM analysis (black/gray) compared to the analysis using the standard BDFM (red/light red).
{\em Lower}: Mean absolute percentage error (MAPE) over time from the 
	LLGM analysis (black) compared to the analysis using the standard BDFM (red).}
 \label{fig:llgmcomp}        
\end{figure}

Some comparison of LLGM and the simple smoothing model~\citep{xi2017bdfm}, referred to as a Bayesian dynamic flow model (BDFM), was made, and one example concerns
counts staying at category Arts \& Entertainment. The discount factor controlling 
adaptability is $\delta=0.9$ for both.   Summaries here focus on forward filtering and one-step ahead forecasting for comparison. 
Both models perform well when the trend is stable---between $8$:$00$ - $19$:$00$, 
the one-step ahead forecasts by both models agree closely with the true data. 
However, BDFM tends to underestimate the rate when the trend is rising and to 
overestimate when the trend is declining, as during $4$:$00$ - $8$:$00$ and $20$:$00$ - $24$:$00$ respectively.
In contrast, LLGM still provides good point-wise prediction. 
Across nearly all flows, LLGM outperforms BDFM in terms of one-step ahead forecast accuracy, illustrated in Fig.~\ref{fig:llgmcomp}.

\section{Model Mapping Analysis \label{sec:dgm} }

\subsection{Daily Fox News Flows Example \label{sec:AdexDGM}}

We now recouple, mapping the retrospective results for the log rate parameters 
$\lambda_{ijt}$ to the DGM. 
The results provide insight into four aspects of flow dynamics: {\em (i)} the 
baseline process $\mu_t$ that characterizes general activity intensity; {\em (ii)} the origin effects $\alpha_{it}$ that relate to activity of outgoing 
flows from node $i$; {\em (iii)} the destination effects $\beta_{jt}$ that relate to  
the attractiveness for incoming flows to node $j$; and {\em (iv)}  the directed 
pairwise affinity effects $\gamma_{ijt}$, for interactions that impact the rate functions governing flows from
nodes $i$ to $j$.

\subsubsection{Baseline level}
\begin{figure}[b!]
    \centering
    \vspace{1em}\ 
     \includegraphics[width=\figsize]{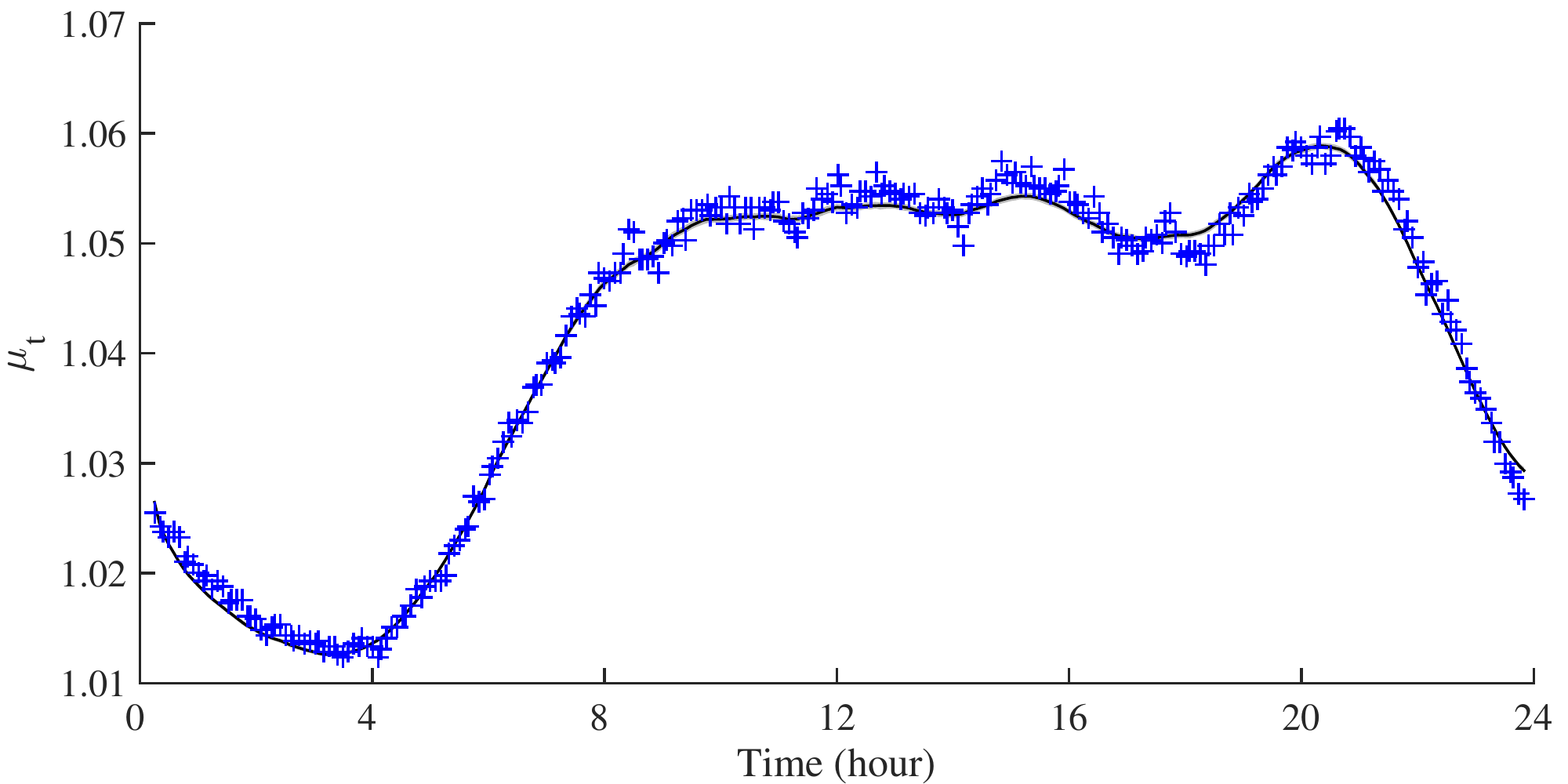} 
      \caption{DGM-based inference on smoothed trajectories of the baseline activity level $\mu_t$ for Fox News data on  September $17$th, $2015$ with a $95\%$ credible interval. The $+$ symbols indicate empirical values from the raw data.}
      \label{fig:Adex_GM_mu}
      \end{figure}

We apply the DGM decomposition to the Fox News data on the network
defined by Adex Category for September $17$th, $2015$.
As expected, the baseline activity level reflects human routine---high traffic
during the day and early evening, and low traffic late at night, as 
shown in Fig.~\ref{fig:Adex_GM_mu}. 

The day starts at midnight when the overall mean intensity is $1.025$.
As users go offline, this overall activity level decreases, reaching a 
minimum at around $4$:$00$.
Then the mean increases until around $8$:$00$, when the trend becomes flat. 
During the day ($8$:$00$ - $16$:$00$), the website maintains a relatively
high level of activity with three small bumps at around $10$:$00$, 
$12$:$30$ and $15$:$00$, which are typical times for work breaks.
There is a slight decreasing trend from $16$:$00$ to $18$:$00$, presumably
as people travel home. 
After dinner time, the trend increases to a peak at around $20$:$00$, and
then declines as people retire. 

\begin{figure}[t!]
    \centering
    \vspace{1em}\ 
\includegraphics[width=5in]{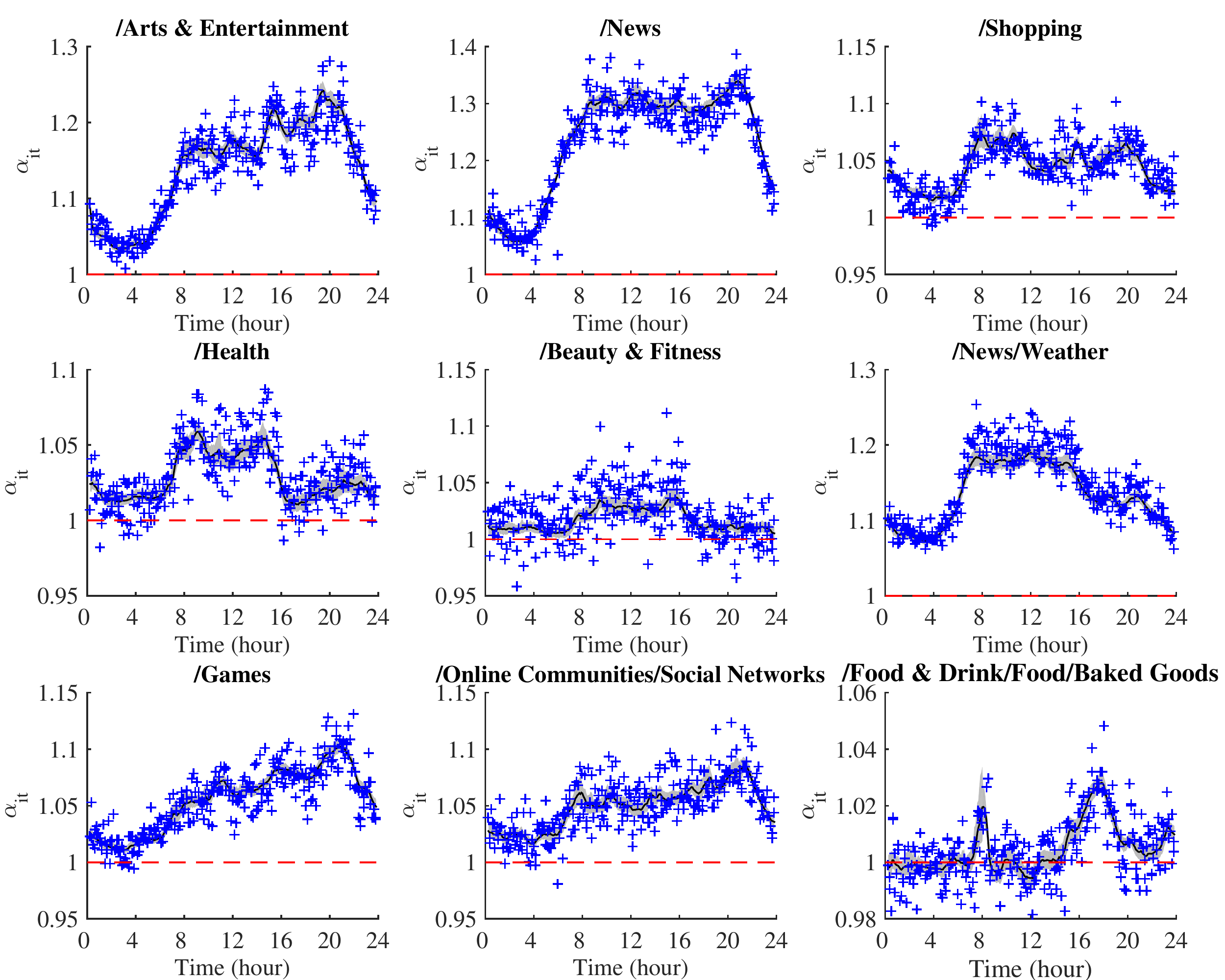}  
\caption{Smoothed trajectories of selected node-specific origin effects $\alpha_{i,1{:}T}$ with $95\%$ CIs.
The $+$ symbols indicate empirical values from the raw data.}
\label{fig:AdexOrigin}
\end{figure}

\subsubsection{Origin and destination effects}
For most categories, the origin and destination trends are  similar so we focus here on the former. Trajectories of origin effects $\alpha_{it}$ exhibit several general patterns,
varying by Adex Category.
Three examples of each appear in Fig.~\ref{fig:AdexOrigin}. 

First, categories such as Arts \& Entertainment, News, and 
Shopping show trends similar to the overall activity level $\mu_t$. 
The origin effect increases to a high level during the early morning,
is steady during the day and early evening, and then declines.
These categories are among the most popular in the network. 

Second, categories such as Health,   Beauty \& Fitness, and 
News/Weather show high activity during the day, but drop to a low 
level in the evening and at night. This is reasonable for weather, since
many people want to know the forecast before leaving for work.  
It also seems reasonable for the other categories, which are oriented towards
a narrow focus that is not entertainment or relaxation.

Third, categories such as Games, Online Communities/Social 
Networks, and Food \& Drink/Food/Baked Goods show a different trend
in origin effect---a pattern that increases from about $4$:$00$ a.m.\ to  
late evening, peaks at around $22$:$00$, and exhibits local peaks
at other times.
Topics on games and social networks pertain to relaxation, which is a 
reasonable evening activity.
The food category has two peaks, in the morning and the late afternoon,
which may reflect people looking at recipes in the morning to plan
grocery purchases, and then later when preparing dinner.

\subsubsection{Affinity effects}
Affinity effects capture interaction between pairs of categories. 
Pairs with strong affinity may indicate that people interested in
one tend to be interested in the other, which would be potentially
valuable in computational advertising.
There are four kinds of affinity effects: {\em (i)} staying at a 
certain category $\gamma_{iit}$; {\em  (ii)} entering the network 
$\gamma_{0jt}; $ {\em (iii)} leaving the network $\gamma_{i0t}$l;  and {\em (iv)} 
moving between two different categories $\gamma_{ijt}$. 
We discuss these separately. For all four kinds of affinities, we show representative
trend patterns, with more extensive examples summarized in Appendix~C.

\paragraph{Self-affinities:} A high self-affinity implies that users tend to stay at that category. 
A trend that shows times of day when people linger is useful to advertisers
since it suggests readers have more leisure time and thus could be tempted to
click on ads.

\begin{figure}[htbp!]
\centering
\includegraphics[width=\figsize]{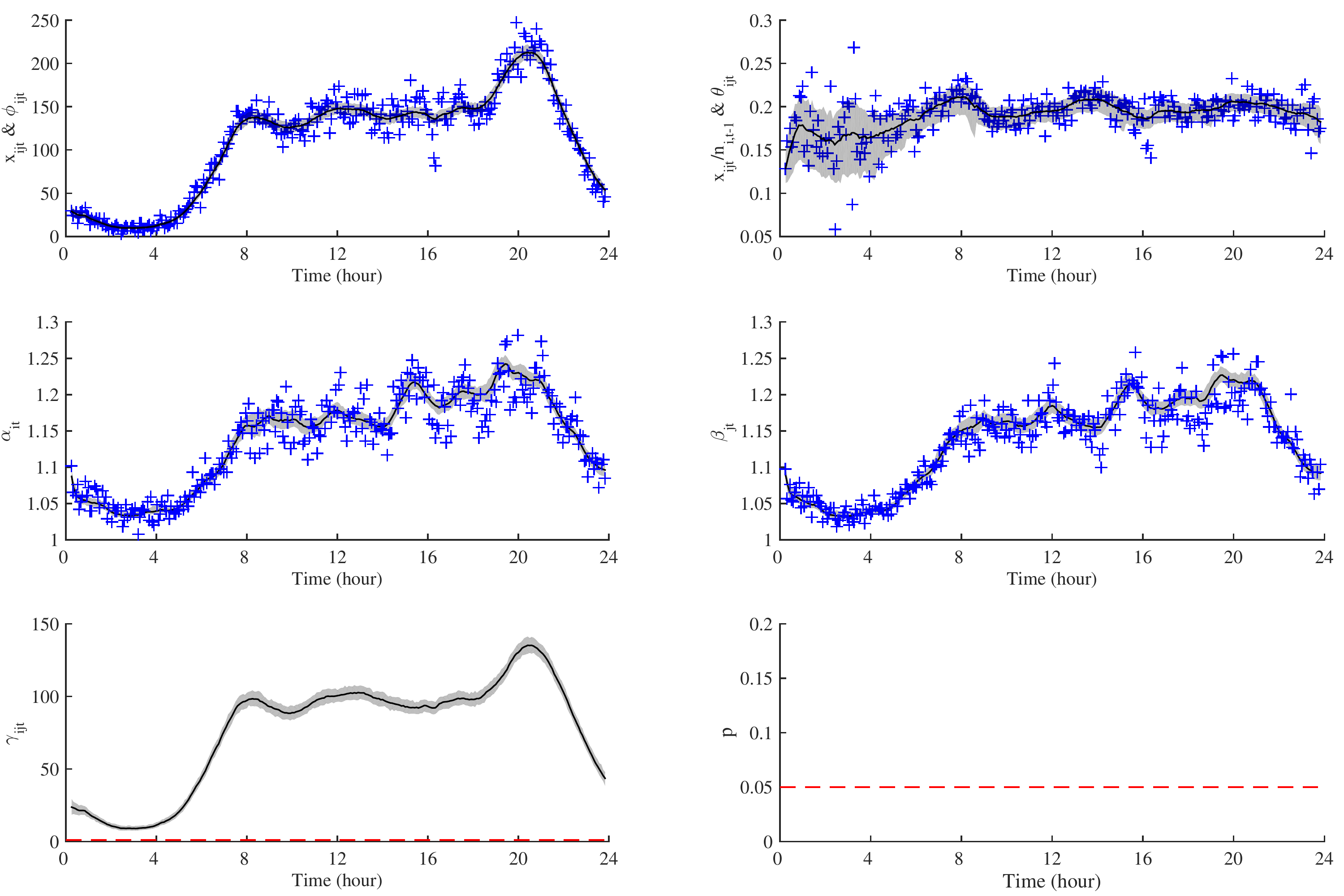}
\caption{Posterior summaries for DGM parameters for transitions staying at node  $i=$ Finance/Investing. 
{\em Upper left:}  Posterior trajectory for the $\phi_{iit}$ with raw counts (crosses). 
{\em Upper right:} Posterior trajectory for the  $\theta_{iit}$ with 
raw frequencies (crosses). 
{\em Center left:}  Posterior trajectory for the origin (outflow) process $\alpha_{it}.$ 
{\em Center right:}  Posterior trajectory for the destination (inflow) process $\beta_{it}.$ 
{\em Lower left:}  Posterior trajectory for the self-affinity
process $\gamma_{iit}$.  
{\em Lower right:} Corresponding trajectories of Bayesian credible values 
assessing support for $\gamma_{iit}$ near 1.}
\label{fig:gmAdex_FinInv} 
\vspace{.4in}
\includegraphics[width=\figsize]{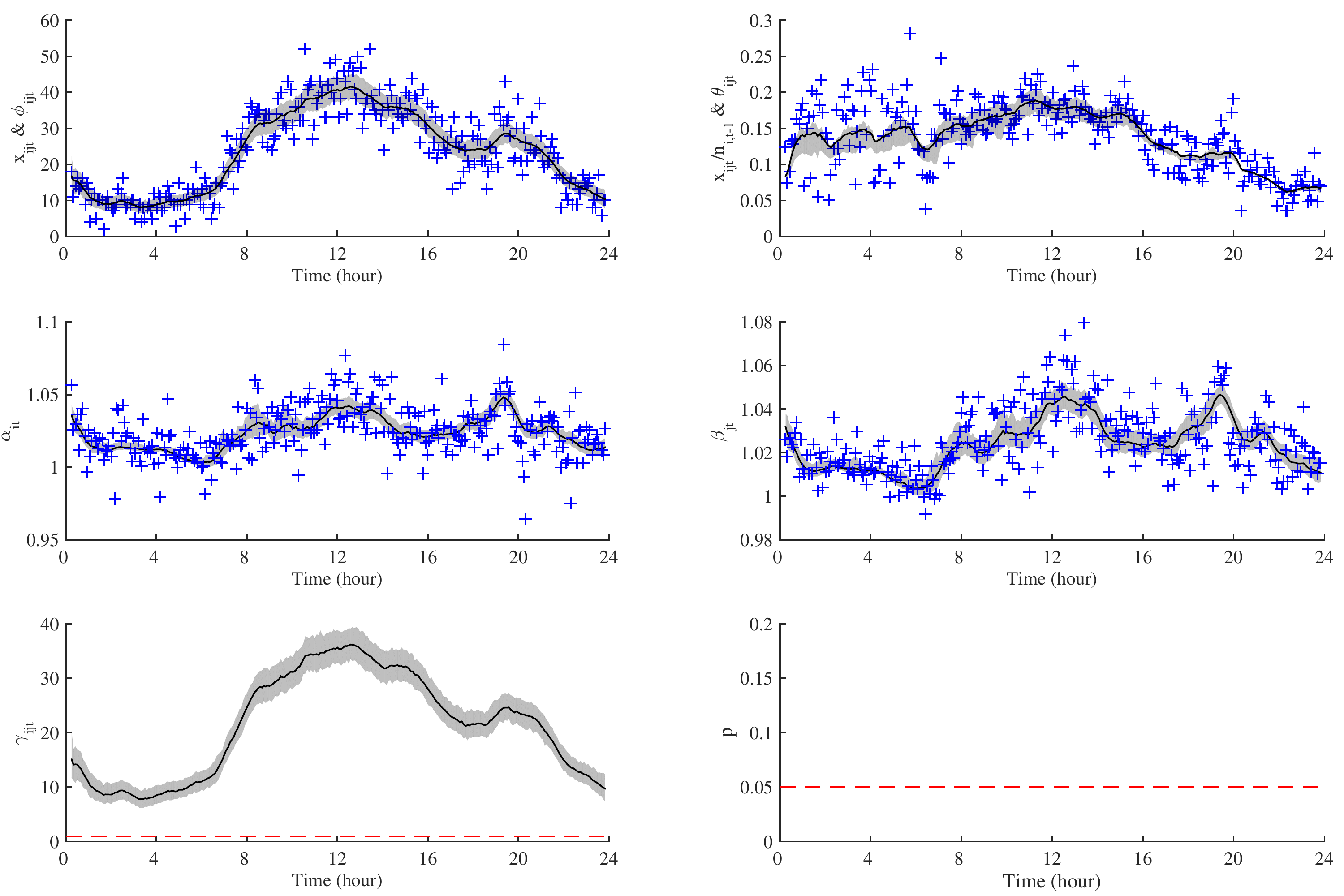} 
\caption{Posterior summaries for DGM parameters for transitions staying at
node $i=$ Computers \& Electronics/Software, with details as in Fig.~\ref{fig:gmAdex_FinInv}.}    
\label{fig:gmAdex_CompAEle}  
\end{figure}

A representative trend pattern features high activity during the business day, and then 
lower level in the evening. 
Such categories include Finance/Investing and 
Computers \& Electronics/Software (Figs.~\ref{fig:gmAdex_FinInv} and 
\ref{fig:gmAdex_CompAEle}, respectively). 
The self-affinity of Finance/Investing has three peaks: one 
around $10$:$00$, one around $15$:$00$, and one 
around $20$:$30$.
Computers \& Electronics/Software is interesting since, for 
most categories, self-affinity drops a bit after $8$:$00$ a.m.\ and is not 
very high at noon, while the affinity of 
Computers \& Electronics/Software increases over the morning and 
peaks at noon.
This trend indicates people spent more time reading related contents compared with other times of the day, and should inform ad buy decisions during those peak times.

\clearpage

\vfill
\vfill

\clearpage

\paragraph{Entering affinities:} A category with high entering affinity draws users 
from outside the Fox News website. 
Ads shown on such categories may be more cost-effective.
The probability $\theta_{0it}$ of users entering the network at node $i$  
is a measure of a category's overall popularity.

\begin{figure}[htbp!]
\centering
\vspace{-1.2em}
\includegraphics[width=\figsize]{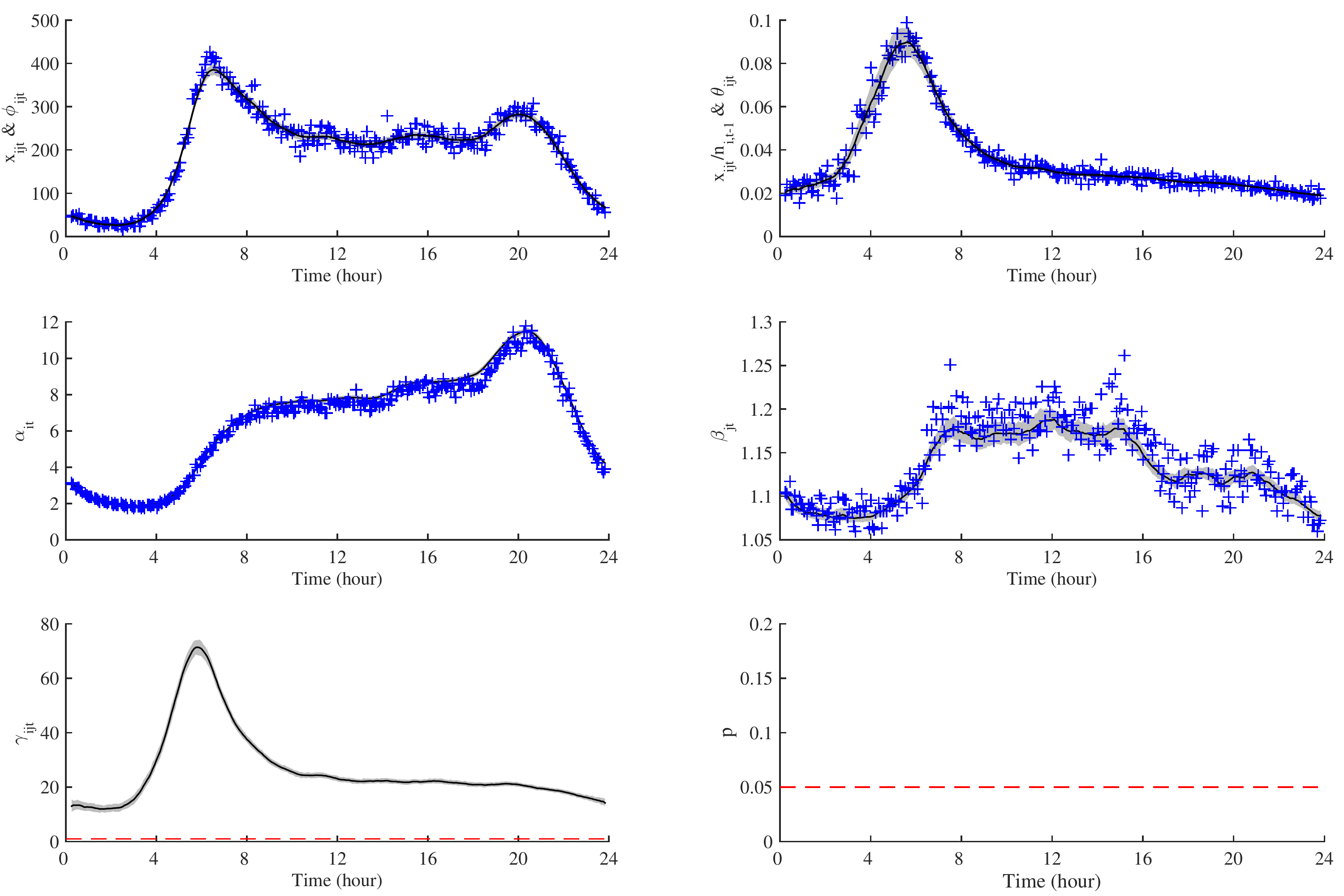} 
\caption{Posterior summaries for DGM parameters for transitions entering node $i=$ News/Weather.  
{\em Upper left:}  Posterior trajectory for the $\phi_{0it}$ with raw counts (crosses). 
{\em Upper right:} Posterior trajectory for the $\theta_{0it}$ with 
raw frequencies (crosses). 
{\em Center left:}  Posterior trajectory for the external origin (outflow)  process $\alpha_{0t}.$ 
{\em Center right:}  Posterior trajectory for the News/Weather destination (inflow)  process $\beta_{it}.$ 
{\em Lower left:}  Posterior trajectory for the News/Weather entering affinity
process $\gamma_{0it}$.  
{\em Lower right:} Corresponding trajectories of Bayesian credible values 
assessing support for $\gamma_{0it}$ near 1.}   
\label{fig:gmAdex_extWth} 
\vspace{.4in}
\includegraphics[width=\figsize]{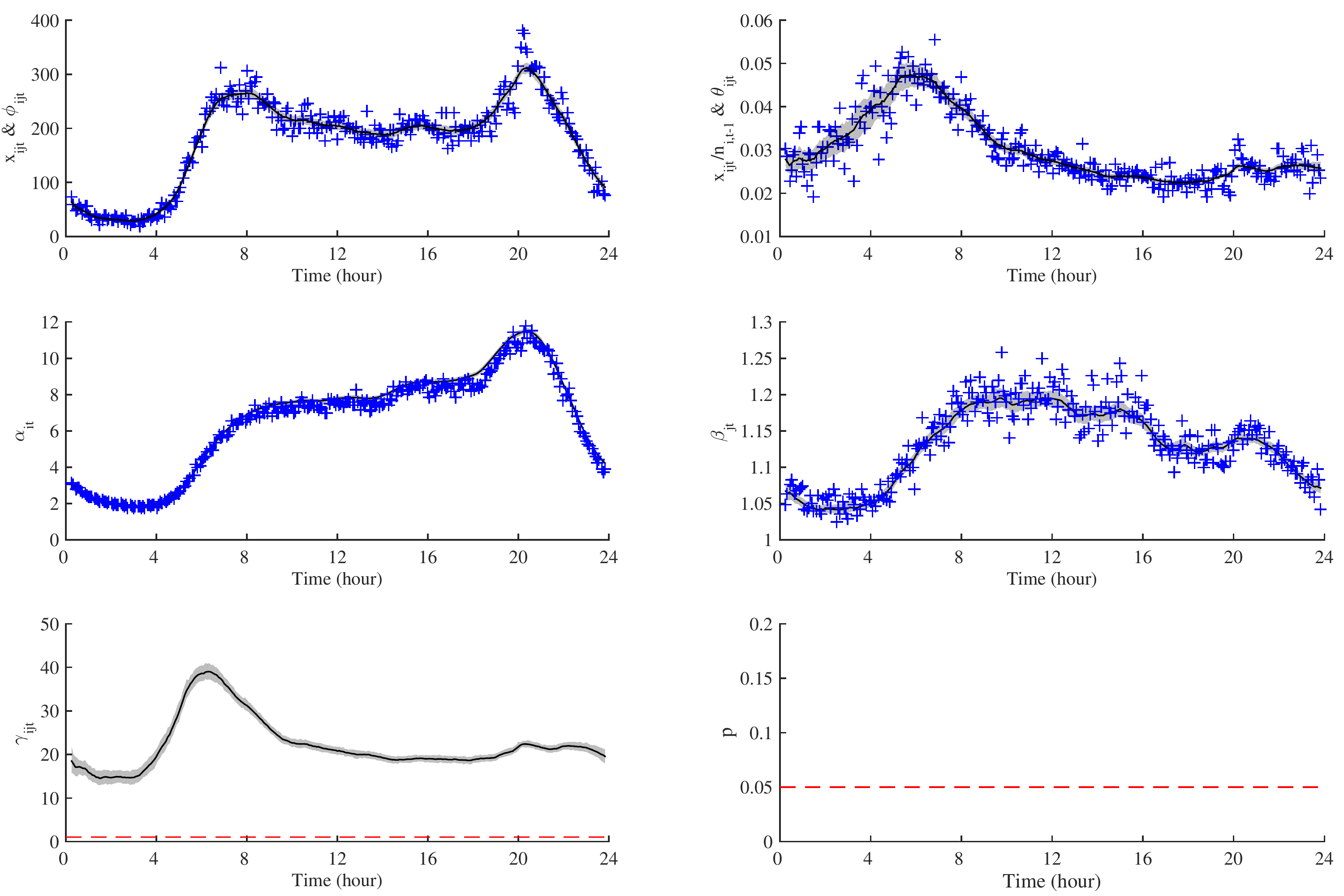}
\caption{Posterior summaries for DGM parameters for transitions entering node $i=$ 
News/Politics with details as in Fig.~\ref{fig:gmAdex_extWth}.}    
\label{fig:gmAdex_extPol} 
\end{figure}

Entering affinities show interesting patterns identified in trajectories. One such 
pattern peaks in the early morning; e.g., flows into News/Weather and News/Politics. 
The peak for both is around $6$:$00$, as in Figs.~\ref{fig:gmAdex_extWth} 
and \ref{fig:gmAdex_extPol}. 
For News/Weather the insight is obvious.
For News/Politics, Sept.\ 17 is the day after a US national political debate, 
which may drive extra interest. That said, this peak pattern is sustained on other 
weekdays, indicating that many people access news sites in the morning.
Advertisers should generally avoid placing ads at this time, since people are 
not in shopping mode.

\paragraph{Exiting affinities:} Categories with high exiting affinities tend to be the last stop
for users navigating the Fox News website. Such categories are the last chance to show an
ad, and if grocery store checkout lines are any guide, a good opportunity for impulse
purchases.

\begin{figure}[htbp!]
\centering
\vspace{-1.2em}
\includegraphics[width=\figsize]{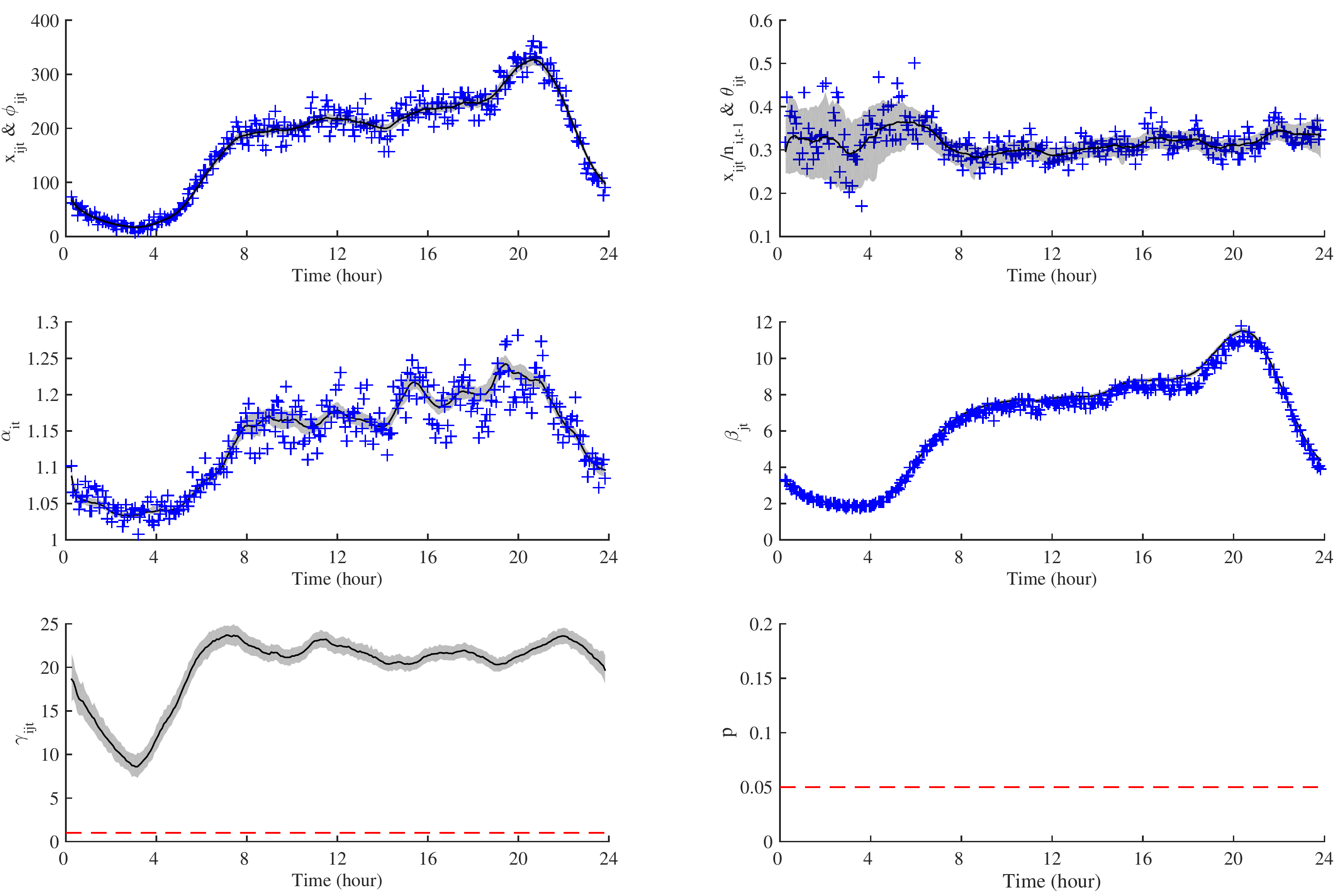} 
\caption{Posterior summaries for DGM parameters for transitions exiting from node $i=$ 
Arts \& Entertainment. 
{\em Upper left:}  Posterior trajectory for the $\phi_{i0t}$ with raw counts (crosses). 
{\em Upper right:} Posterior trajectory for the $\theta_{i0t}$ with 
raw frequencies (crosses). 
{\em Center left:}  Posterior trajectory for the Arts \& Entertainment origin (outflow)  process $\alpha_{it}.$ 
{\em Center right:}  Posterior trajectory for the external destination (inflow)  process $\beta_{0t}.$ 
{\em Lower left:}  Posterior trajectory for the Arts \& Entertainment leaving affinity process $\gamma_{i0t}$.  
{\em Lower right:} Corresponding trajectories of Bayesian credible values 
assessing support for $\gamma_{i0t}$ near 1.}    
\label{fig:gmAdex_Artext} 
\vspace{0.4in}
\includegraphics[width=\figsize]{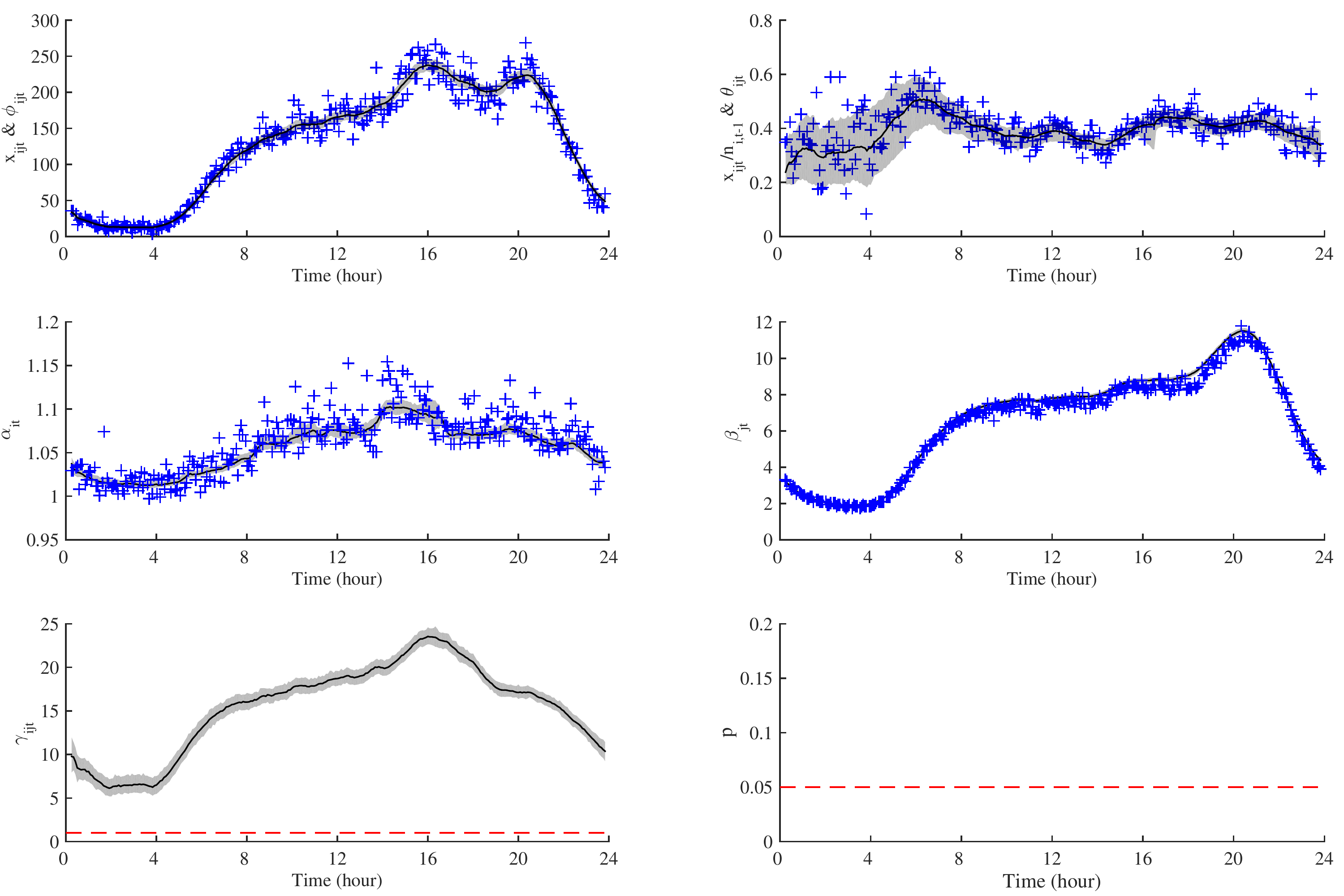} 
\caption{Posterior summaries for DGM parameters for transitions exiting from node $i=$ 
Food \& Drink/Cooking \& Recipes with details as in Fig.~\ref{fig:gmAdex_Artext}.}   
\label{fig:gmAdex_Cookext}
\end{figure}

One interesting trend increases over the morning, peaks at about $8$:$00$ and then 
stays stable with high intensity until evening.
Often, these categories are the only categories visited by a user. 
Examples include Arts \& Entertainment and Food \& Drink/Cooking 
\& Recipes (Figs.~\ref{fig:gmAdex_Artext} and \ref{fig:gmAdex_Cookext}). 
Though the affinity intensity is stable, there is an interpretable bump 
during $16$:$00$ - $18$:$00$ for Food \& Drink/Cooking \& Recipes
which is probably related to dinner preparation.

\paragraph{Distinct node pair affinities:} 
High levels of affinity between distinct category pairs indicate interaction, which
could reasonably influence advertising strategy. 

In all patterns of such affinities,
the one with a bump is the kind in which we are most interested.
The bump indicates that users are
only active in moving between those categories in a short window, while they remain inactive 
during the rest of the day, and thus this short window is the best time that related ads should 
be displayed. Examples include Online Games to Video Games and Home \& Garden to Reference/General Reference/How 
to DIY \& Expert Content.

\begin{figure}[htbp!]
\centering  
\vspace{-1.2em}
 \includegraphics[width=\figsize]{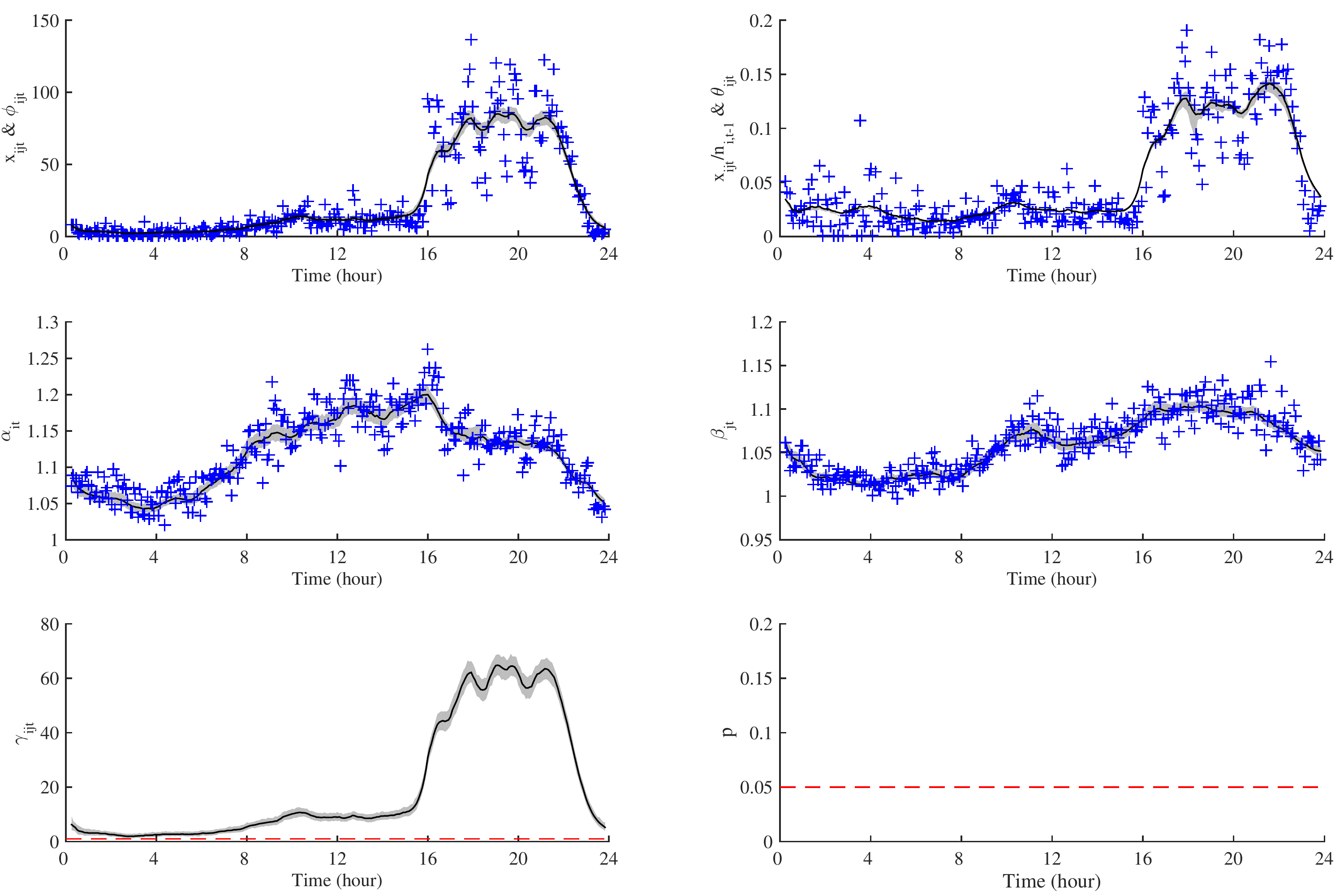} 
\caption{Posterior summaries for DGM parameters for transitions from node $i=$
Games/Online Games $\rightarrow j=$ Games/Computer \& Video Games.  
{\em Upper left:}  Posterior trajectory for the $\phi_{ijt}$ with raw counts (crosses). 
{\em Upper right:} Posterior trajectory for the $\theta_{ijt}$ with 
raw frequencies (crosses). 
{\em Center left:}  Posterior trajectory for the Games/Online Games origin (outflow)  process $\alpha_{it}.$ 
{\em Center right:} Posterior trajectory for the News/Weather destination (inflow)  process $\beta_{jt}.$ 
{\em Lower left:}  Posterior trajectory for the Games/Online Games : Games/Computer \& Video Games affinity process $\gamma_{ijt}$.  
{\em Lower right:} Corresponding trajectories of Bayesian credible values 
assessing support for $\gamma_{ijt}$ near 1.}    
\label{fig:gmAdex_GamesVideo} 
\vspace{0.4in}
\includegraphics[width=\figsize]{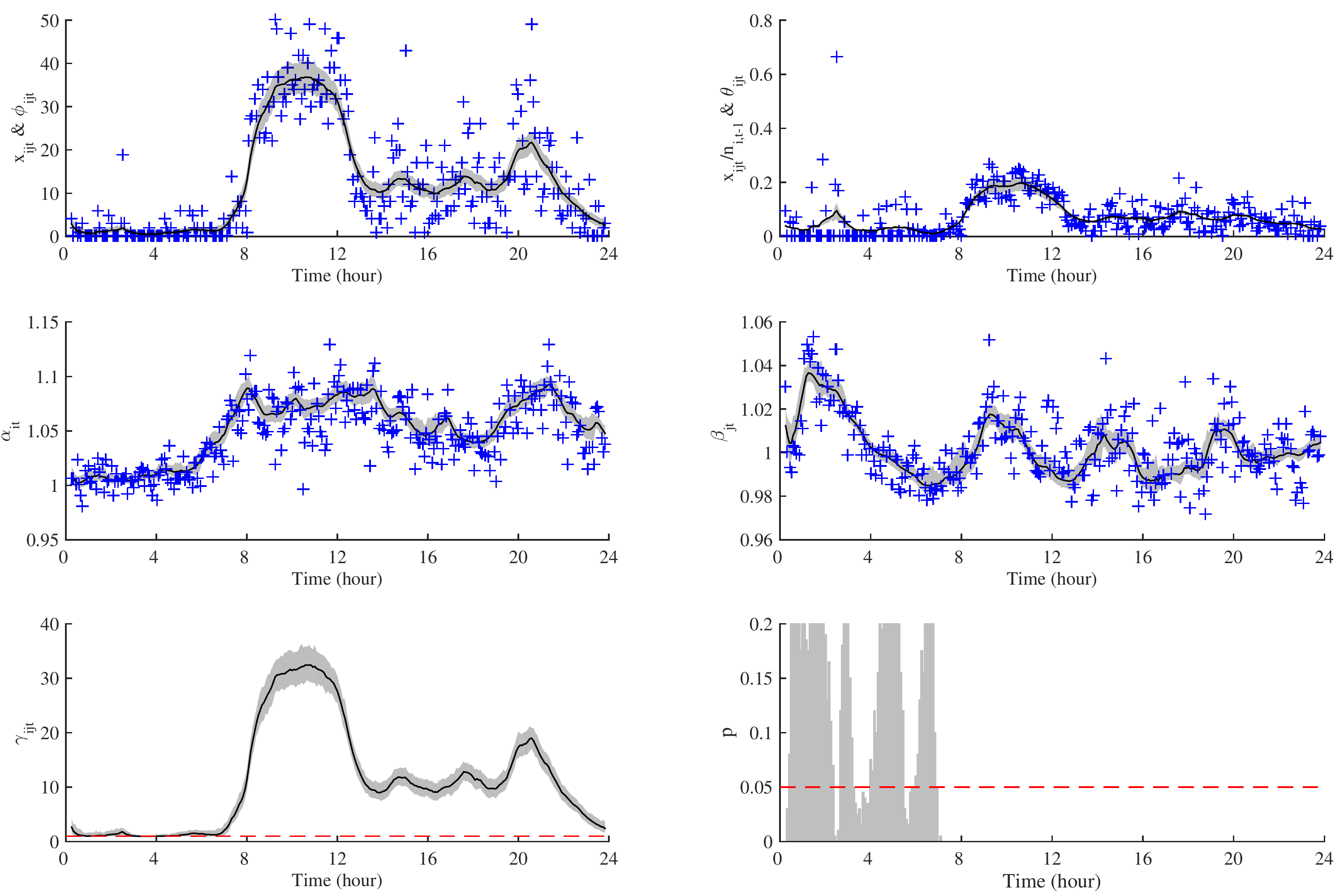} 
\caption{Posterior summaries for DGM parameters for transitions from node $i=$
Home \& Garden $\rightarrow j=$ Reference/General Reference/How to DIY \& Expert Content with details as in Fig.~\ref{fig:gmAdex_GamesVideo}.}   
\label{fig:gmAdex_HomeDIY} 
\end{figure}

The affinity from Online Games to Video Games
(Fig.~\ref{fig:gmAdex_GamesVideo}) has a bump in the evening and at 
night ($16$:$00$ - $24$:$00$) during which the average is six times 
higher than the usual intensity level.
This strongly indicates that users who read about online games are also 
interested in computer and video games during this period. 
 
The affinity from Home \& Garden to Reference/General Reference/How 
to DIY \& Expert Content (Fig.~\ref{fig:gmAdex_HomeDIY}) has a bump 
around $8$:$00$ - $12$:$00$. 
Moreover, both of the origin effect of Home \& Garden and the 
destination effect of Reference/General Reference/How to DIY \& 
Expert Content are low, while the affinity effect between them is 
large, which indicates strong interaction. 
Obviously, people plan home projects in the morning, and seek information
on how to implement them.

\clearpage

\clearpage
\clearpage

\clearpage
\clearpage

\section{Closing Comments \label{sec:closing}} 

The case study demonstrates the value of this new class of dynamic network flow models on a large network. 
The specific special case of DGLMs adopted---the local linear growth model (LLGM)---is just one example of the broader class, but the analysis examples highlight the ability of this family to characterize and adapt to quite heterogeneous patterns of change over time in latent 
flow patterns. An important element of the analysis is that full Bayesian inference, based on computationally efficient retrospective sampling analysis, defines not only point estimates of trajectories of key dynamic parameters, but also uncertainties about them from posterior samples. 

A  critical  component of the analysis is the strategy of decoupling/recoupling.
This has two aspects.  First, the individual univariate DGLMs for dynamic Poisson flows are learned by forward filtering updates at each time step,  and then for all within-network nodes these are theoretically recoupled to define inferences on the dynamic transition probability vectors in the inherent  conditional multinomial distributions governing flows, so inferring node dependencies. 
The decoupling/recoupling strategy also provides an ideal structure to perform parallel computation, enabling scalable and efficient analysis for increasingly large dynamic networks. Using a computer with $K$ cores, the computational demands for a series of length $T$ on a network with $I$ nodes  scales only as $\mathcal{O}(I^2T/K)$. A 2018 MATLAB implementation running on a standard laptop (2.3GHz cpu, 16Gb memory) took $52$ minutes to run our analysis with $T = 286$, $I = 2{,}208$.

The second aspect of recoupling analysis is the use of the set of DGLM-based models as an emulator for a multivariate dynamic gravity model that is explicitly parametrized in terms of network-wide, node-specific and node pair interaction effects, all potentially time-varying.  The Fox News web study shows a number of examples of the use of this mapping to uncover---again with full Bayesian posterior uncertainty measures---interpretable patterns in time trajectories of node main effects (impacting both incoming and outgoing traffic)  and node-node pair interactions (in terms of affinities of one node for another that impact traffic flows between them).  We have noted the potential for such information to be exploited by on-line advertisers in the case study, and formal inferences will be of interest in other applications.   

The DGM mapping also enables investigation of network grouping structures over time by clustering nodes using the dynamic node-node pair affinities, with potential for deeper future study. With the overall network intensity and node main effects removed by DGM, the affinity between a pair of nodes is an inherently interesting measure  of the \lq\lq closeness'' of nodes; at one level, it can be regarded as a coherent statistical summary of node-node   relationship and their patterns over time (and much preferred to any raw data summary).  We note, in addition, that the node-specific main effects for inflows can be regarded as statistical quantities of node importance or centrality; inflows and outflows  together reflect the scale of importance of a node in terms of numbers and flow intensities with neighbors in the network. Traditional summary network topology measures may be of interest in other contexts, but for dynamic network traffic analysis these model-derived quantities are foremost, and their inferred trajectories over time are of primary interest. 

The case study exploits and highlights the adaptability and utility of the local trend LLGM class of models as a key driver of the overall decouple/recouple and emulation analysis.  This is, of course, just one special subclass of the full class of DGLMs, and other forms are applicable in other areas. 
Future studies with available covariate information might include, for example,   internet traffic studies with known interventions,  dynamic traffic flow with geographical and structural information, or brain network data with known or hypothesized connectivity information. 
Such studies can be expected to exploit more general DGLMs that include node-specific covariates or dummy variables representing intervention effects (e.g., for known ad placements in e-commerce examples, or  network structure changes in others).  This work offers benefits in other applications through the flexibility to customize the DGLMs to the context.   The overall model framework and approach needs only customization of details in the specification of the state-space elements $\F_t,\G_t$  (which will in some extensions be specific to pairs of nodes as well as time-varying), as the analysis described here and illustrated in the case study applies generally to the full DGLM class.

\clearpage

\section*{Appendix} 

\subsection*{Appendix A:  Recouple Mapping to Dynamic Gravity Model }

The mapping from the DGLM flow model to DGM parameter processes of Section~\ref{sec:mapptoDGM} follows developments in~\cite{xi2017bdfm}, summarized as follows.  We work on logged DGM parameters  $h_{t}=\log(\mu _t)$,  $a_{it}=\log(\alpha_{it})$, $b_{jt}=\log(\beta_{jt})$ and 
$g_{ijt}=\log(\gamma_{ijt}).$     To ensure identification and a one-one mapping between the models,   the traditional zero-sum constraint is adopted 
for the main effects and interactions at each time.   Then, given the full set of $\phi_{ijt}$ values in the DGLM flow model, define 
 $f_{ijt}=\log(\phi_{ijt})$ for each $i=1{:}I, j=0{:}I,$ at each time $t=1{:}T.$    The recoupling step is then implemented by computing DGM parameters as below.  In the notation, subscript $+$ indicates summation over the relevant index $i$ or $j.$ For each $t,$ 
\begin{itemize} \setlength\itemsep{0pt}
\item compute $\mu_t=\exp(h_t)$ where $h_t =
  f_{++ t}/I(I+1)$;  then, 
\item for each $i=1{:}I,$ compute
$\alpha_{it}=\exp(a_{it})$ for $a_{it} = f_{i+ t}/(I+1) -
h_t$;  then, 
\item for each $j=0{:}I,$ compute the destination node main effect
  $\beta_{jt}=\exp(b_{jt})$ for $b_{jt} = f_{+ jt}/I - h_t$;  finally, 
\item for each $i=1{:}I$ and $j=0{:}I,$ compute the affinity (interaction) effect 
  $\gamma_{ijt}=\exp(g_{ijt})$ where $g_{it} =f_{ijt} - h_t - a_{it} - b_{jt}.$
\end{itemize}
This is applied to all simulated  $\phi_{ijt}$ values from the posterior analysis under
the DGLMs to create implied  posteriors for the DGM parameter processes.

\subsection*{Appendix B:  Bayesian Analysis of Poisson DGLMs}

Analysis is based on sequential Bayesian computation that combines variational Bayes approximation with linear 
Bayes updates~\citep{WestHarrison1997,hartigan1969linear,goldstein1976bayesian}, along with retrospective sampling of posteriors for state vector trajectories, extending earlier algorithms for DGLM  analysis. 

As in Section~\ref{sec:dglm}, we focus on one node pair but ignore the node indices, so that we have a Poisson DGLM for $x_t$ with mean $\phi_t$ where 
 $\lambda_t = \log(\phi_t)$ is defined via the underlying linear, state-space, Markov model of \eqno{DGLMevolution}.

\subsubsection*{Sequential analysis: Forward filtering and learning} 

At time $t=0$, specify a prior mean vector $\m_0$ and variance matrix $\C_0$ for the pre-initial state vector, denoted by 
$\bth_0 \sim [\m_0,\C_0].$   Then, over every future time point $t>0,$ the evolution over $t-1$ to $t$, prediction of $x_t$ from time $t-1$ and posterior update based on observing $x_t$ follows the standard evolve/predict/update cycle of Bayesian state-space model analyses. In DGLMs, we use two approximations in the cycle. First, as the evolution noise distribution is specified only in terms of first and second order moments, the modeler is free to constrain implied state and predictive distributions to chosen forms;  DGLMs use the variational Bayes concept to constraint to conjugate forms enabling fast and efficient computation, as well as defining predictive distributions of contextually appropriate  (for count time series) negative binomial forms.  This is coupled with the use of decision-theoretic linear Bayes approximations to feed back data information in the prior-posterior update (filtering) step at each time step, appropriately conditioning the mean vector and variance matrix for the state vector as new data is processed. Details are summarized below. 
 
\begin{enumerate} 
\item  
At time $t-1$, given all the previous data and information $\cD_{t-1}$  the mean vector and variance matrix of the posterior for $\bth_{t-1}$ are available as 
$\bth_{t-1}|\cD_{t-1} \sim [\bm{m}_{t-1},\bm{C}_{t-1}]$.
\item 
By the state evolution equation, the implied time $t-1$ prior for $\bth_t$ has moments 
$ \bth_{t}|\cD_{t-1} \sim [\bm{a}_{t},\bm{R}_{t}]$,
where   $\bm{a}_t=\G_t\bm{m}_{t-1}$
and   $\bm{R}_t=\G_t\bm{C}_{t-1}\G'_t+\W_t$.
\item This implies that $\lambda_t = \log \phi_t$ has prior mean and variance given by 
 $f_t = \F'_t\bm{a}_t$ and $q_t = \F'_t\bm{R}_t\F_t$ respectively.
 
\item The implied prior for the latent Poisson rate is constrained to a conjugate gamma form based on the above information---a variational Bayes decision and constraint.  That is, the modeling choice is made to specify 
$\phi_t|\cD_{t-1} \sim Ga(r_{t},c_{t})$ with defining parameters consistent with the prior information, i.e., consistent with the moment constraints 
about on $\lambda_t = \log \phi_t.$   Matching these moments to the gamma prior implies that $r_{t},c_{t}$ are given as solutions to 
the equations $f_t  = \gamma(r_t)-\log c_t$ and $q_t= \dot{\gamma}(r_t),$ respectively, where $\gamma(\cdot)$ is the digamma function and $\dot{\gamma}(\cdot)$ is 
the trigamma function. 
 These equations are easily solved numerically (most efficiently using the Newton-Raphson method) to give the values of $r_t,c_t.$   

\item  Forecasting $x_t$ from time $t-1,$ the conjugate Poisson/gamma structure implies a negative binomial predictive distribution $p(x_t|\cD_{t-1}).$ 
\item  On observing $x_t$ at time $t,$ the implied posterior for the latent Poisson rate is the conjugate
form  $\phi_t|\cD_t \sim Ga( r_t+x_t,c_t+m_{t})$ (where $m_t$ is the relevant occupancy correction factor).  For the natural parameter $\lambda_t = \log \phi_t,$ this implies posterior moments 
$\lambda_t|\cD_t \sim [f^*_t,q^*_t]$ given by 
$$f^*_t  = \gamma(r_t+x_t)-\log(c_t+m_{it})\quad\mbox{and}\quad q^*_t = \dot{\gamma}(r_t+x_t)$$
and these are trivially calculated. 
\item  Using linear Bayes decision theory arguments,  the posterior mean vector and variance matrix for the state vector $\bth_t$ are conditioned on the 
new information $x_t$ via the predictor-corrector forms that adjust the prior moments based on forecast accuracy. Specifically, the time $t-1$ to $t$ posterior update of moments $[\m_\ast, \C_\ast]$ required to complete the time $t$ filtering steps are given by
$$
\m_t = \a_t+\A_t(f^*_t-f_t)/q_t \quad\mbox{and}\quad 
\C_t = \R_t-\A_t\A_t'(q_t-q_t^*)
$$
where $\A_t$ is the adaptive coefficient vector $\A_t=\R_t\F_t/q_t.$ 

\end{enumerate}

\subsubsection*{Discount specification of evolution variance matrices} 
 
Specification of the evolution variance matrix $\W_t$ at each time uses the standard single-discount factor approach in which 
$\W_t = (\G_t\C_{t-1}\G_t') (1-\delta)/\delta$.  This corresponds to the time $t-1$ posterior variance matrix $V(\bth_{t-1}|\cD_{t-1}) = \C_{t-1}$ evolving to the
prior variance matrix $V(\bth_t|\cD_{t-1}) = \R_{t-1}= \G_t\C_{t-1}\G_t' + \W_t = (\G_t\C_{t-1}\G_t')/\delta. $ That is, following the deterministic component (defined by $\G_t$) of the state evolution,  the stochastic innovation term $\bom_t$ in the evolution increases uncertainties about the state by \lq\lq discounting'' historical information at a rate defined by $\delta.$  The discount factor 
$\delta$ is a tuning parameter to be chosen. See, for example, \citet[Chapter 6]{WestHarrison1997}, \citet[Section 4.3.6]{PradoWest2010}.

\subsubsection*{Retrospective analysis}

The above analysis provides for sequential learning, i.e., forward filtering to process data as it arrives and sequentially update prior and posterior mean and variance matrices for the state vector $\bth_t$ over time.  At any time $t,$ this enables inference on the current state and forecasts of coming data. For most network studies this is most relevant to online learning and monitoring of flows.  Then, having processed data up to any time $T$,  a key interest is in looking back over time to update inferences on the historical trajectories of state vectors, and any functions of them of interest. This is called retrospective analysis and 
can be best addressed in terms of simulation of posterior distributions over the full past history of states.  This is enabled using DGLM extensions of the standard \lq\lq backward sampling'' algorithm for conditionally Gaussian, dynamic linear models.   That is, exploiting the retrospective extrapolation of posterior mean vectors and variance matrices, and adopting the variational Bayes concept again to constrain the implicit backward innovations to multivariate normal distributions with moments defined by the linear Bayes retrospection, we impute and simulate historical {\em trajectories} of sets of states $\bth_1,\ldots,\bth_T$ by recursing backwards in time as follows.  

\begin{itemize} 
\item At time $t=T$, sample the approximating normal posterior  $\bth_t|\cD_T \sim N(\m_T,\C_T).$ 
\item Recurse back over times $t=T-1,T-2, \ldots,1,$ at each stage sampling the variational Bayes normal approximation to $p(\bth_t|\bth_{t+1},\cD_T).$
\item At $t=1,$ save the sampled trajectory of states. 
\item Repeat to generate a random sample of trajectories.
\end{itemize} 
From a Monte Carlo sample of trajectories, we can then directly map to any functions of the state vectors for inference. Centrally, this includes mapping to the
sampled Poisson rate parameters and then to the network-wide, node-specific and node-node interaction parameter processes of the dynamic gravity models.

In terms of implied computation, the retrospective theory simplifies considerably in the single discount DGLM.   Summary computations for simulation at time $t$ of the state vector $\bth_t$ from the implied conditional normal for $\bth_t | \bth_{t+1},\cD_T$ has mean vector and variance matrix given by 
simplified versions of the general expressions in dynamic linear models~\citep[e.g.][Sections 4.3.5 and 4.3.6]{PradoWest2010}. In detail, the conditional moments are simply
$$ E(\bth_t|\bth_{t+1},\cD_T) =   (1-\delta)\m_t+\delta \G^{-1}\bth_{t+1}\quad\mbox{and}\quad 
    E(\bth_t|\bth_{t+1},\cD_T) =  (1-\delta)\C_t, $$ 
    and the implied approximate normal distribution is trivially simulated.

\subsection*{Appendix C: More Examples of Affinity Effects }

\subsubsection*{Self-affinities}
\begin{figure}[htbp!]
\centering
\includegraphics[width=\figsize]{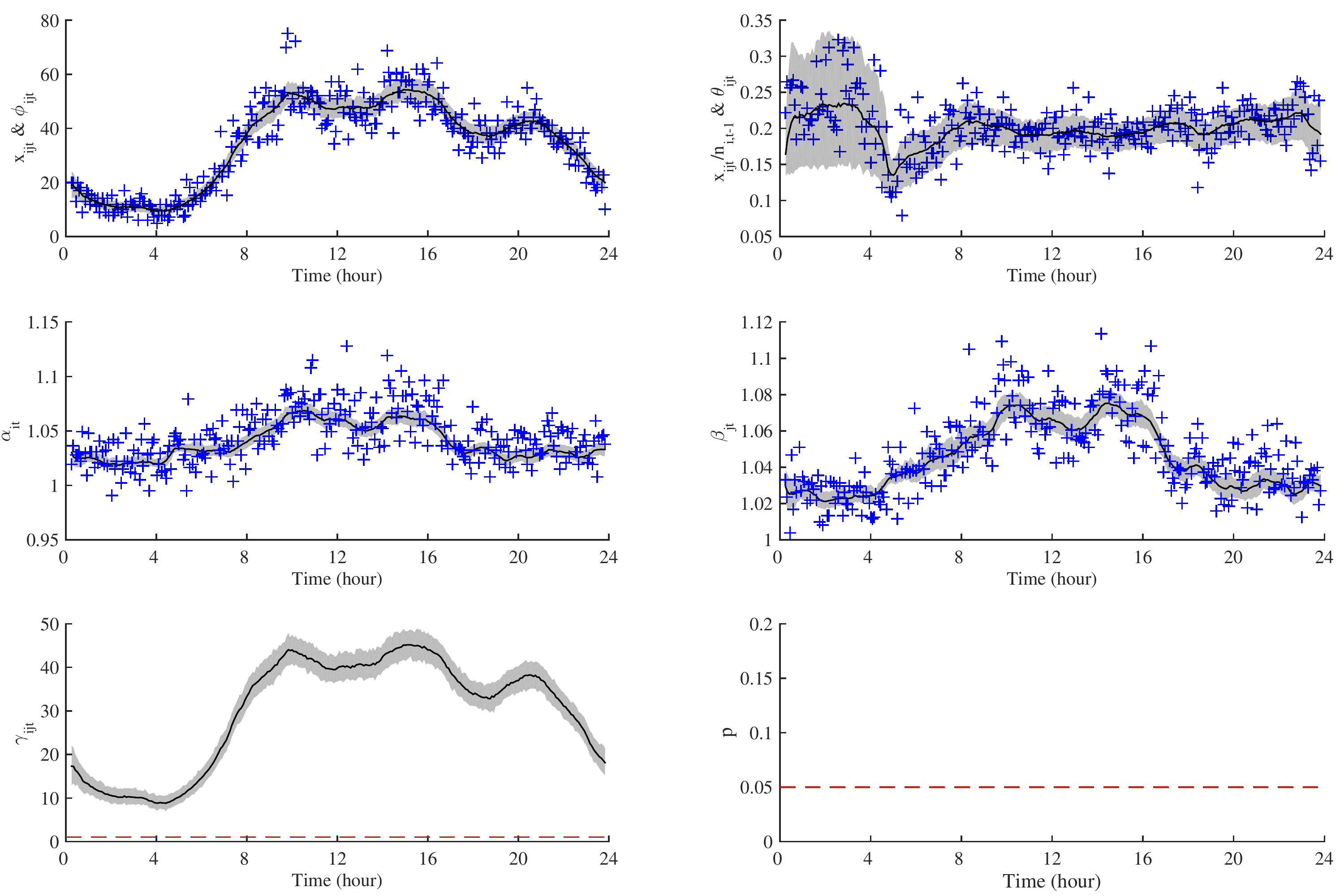} 
\caption{Posterior summaries for DGM parameters for transitions staying at node $i=$ Arts \& Entertainment with details as in Fig.~\ref{fig:gmAdex_FinInv}.}   
\label{fig:gmAdex_ArtEnt}  
\bigskip\bigskip
\includegraphics[width=\figsize]{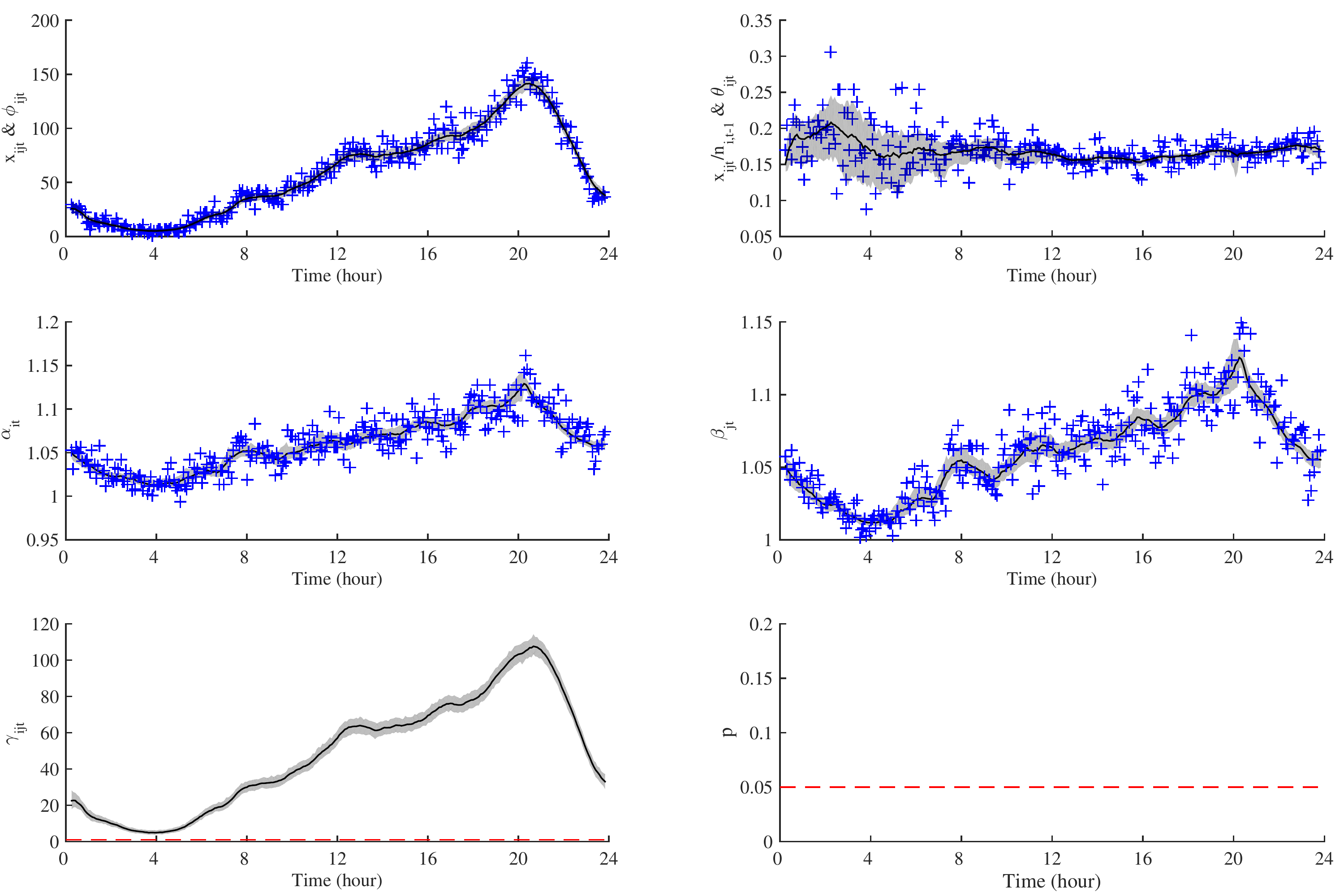} 
\caption{Posterior summaries for DGM parameters for transitions staying at node $i=$ Arts \& Entertainment/TV \& Video,  with details as in Fig.~\ref{fig:gmAdex_FinInv}.}    
\label{fig:gmAdex_TVVideo}  
\end{figure}

Another pattern in self-affinities that is  common across    nodes resembles the overall activity $\mu_t$; category Arts \& Entertainment is a good example
(Fig.~\ref{fig:gmAdex_ArtEnt}). 
Levels are high during the day and evening ($8$:$00$ - $20$:$00$) with a 
large bump around $20$:$00$ p.m. 
We also see a pattern that increases throughout the day and peaks at night; 
the category Arts \& Entertainment/TV \& Video is one of the few examples (Fig.~\ref{fig:gmAdex_TVVideo}).

\subsubsection*{Entering affinities}
\begin{figure}[htbp!]
\centering
\includegraphics[width=\figsize]{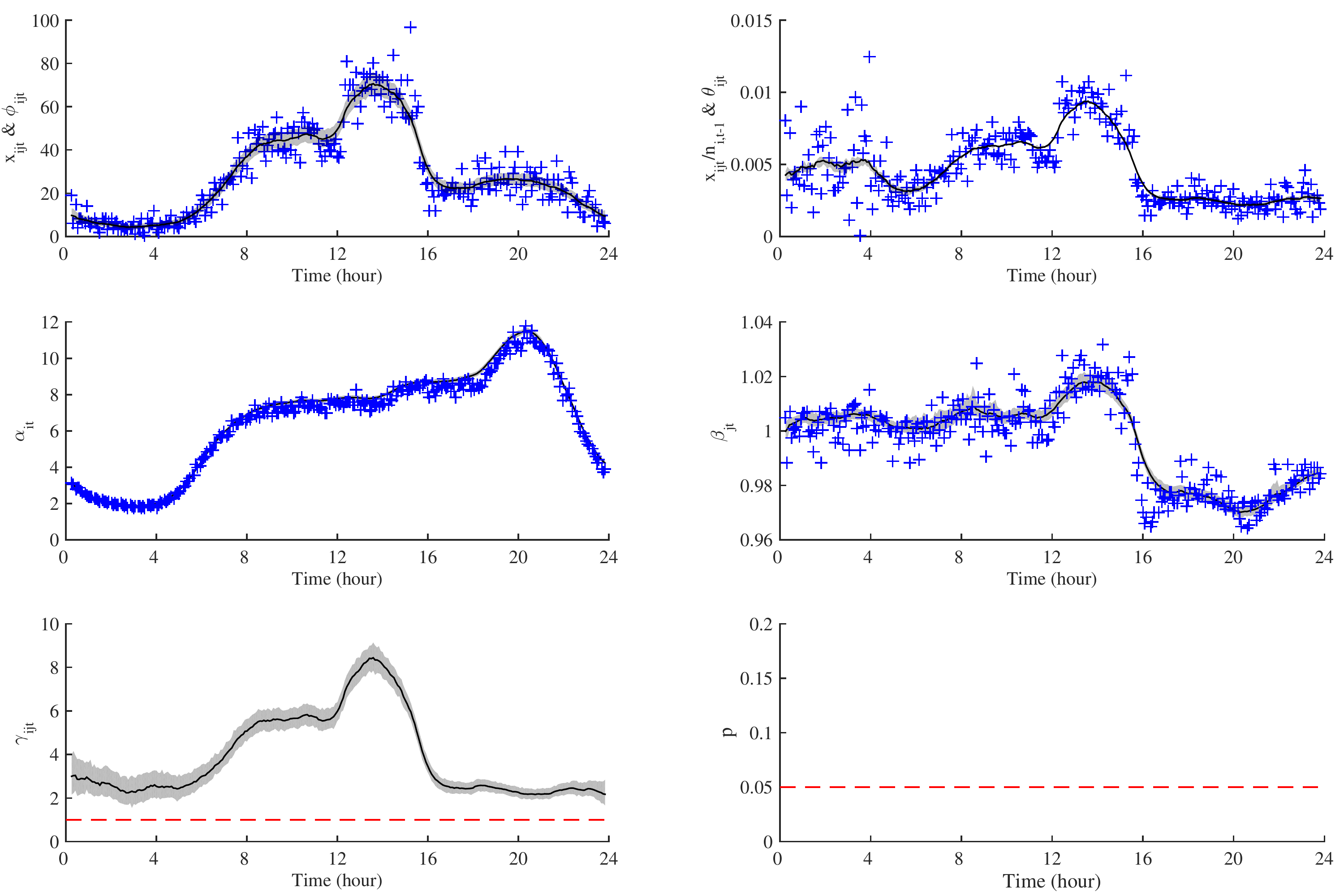} 
\caption{Posterior summaries for DGM parameters for transitions entering node $i=$ Health/Pharmacy/Drugs \& Medications with details as in Fig.~\ref{fig:gmAdex_extWth}. An interesting bump is noted in the afternoon (about $12$:$00$ - $16$:$00$).}    
\label{fig:gmAdex_extDrug}
\bigskip\bigskip
\includegraphics[width=\figsize]{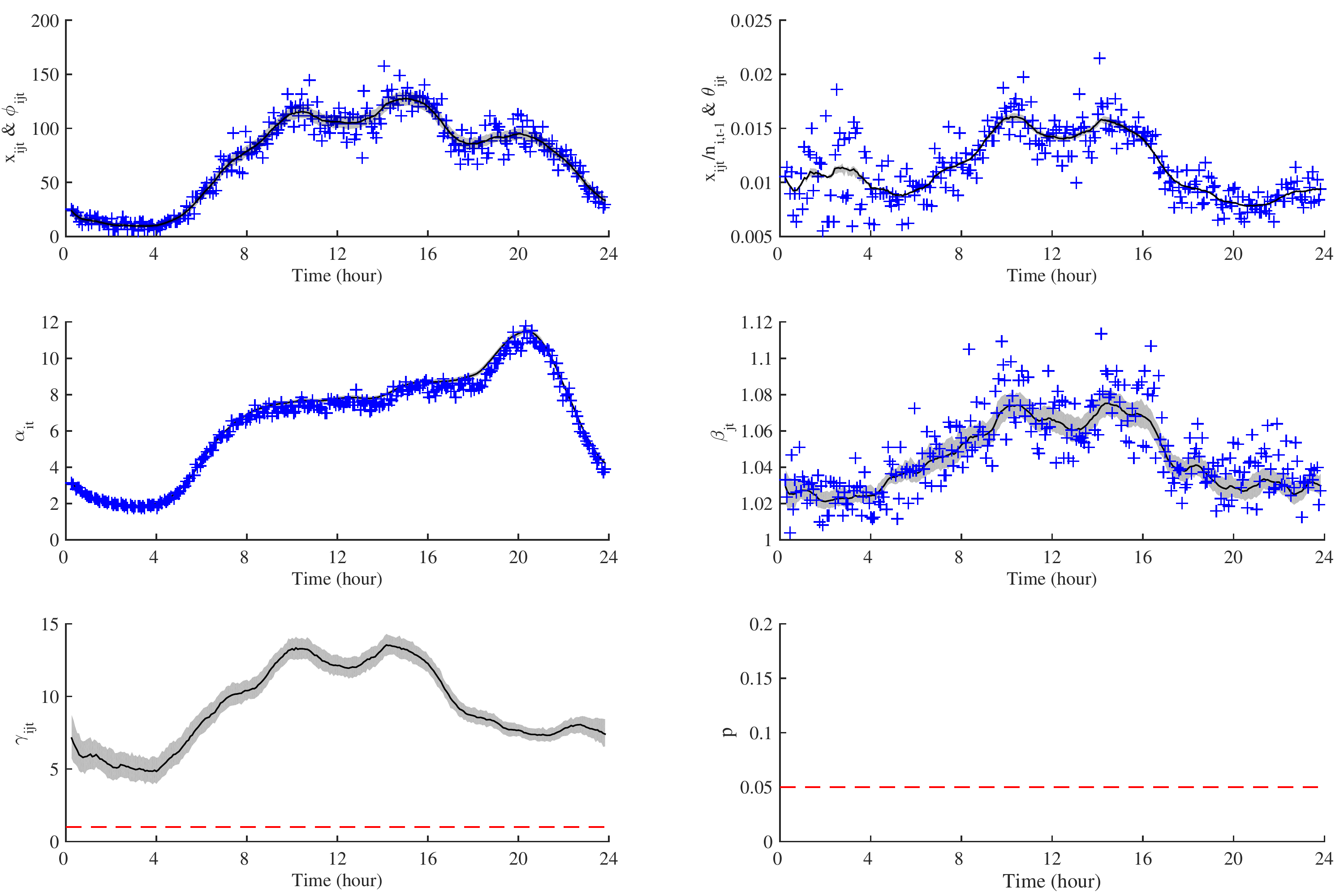} 
\caption{Posterior summaries for DGM parameters for transitions entering node $i=$ 
Finance/Investing with details as in Fig.~\ref{fig:gmAdex_extWth}. 
Note a maintained high level during the day ($8$:$00$ - $16$:$00$).}    
\label{fig:gmAdex_extFinInv}
\end{figure}

Beyond the examples in Section~\ref{sec:AdexDGM}, another pattern in entering affinities, exemplified by  the 
categories  Health/Pharm\-acy/\-Drugs \& Medications and Finance/Investing
(Figs.~\ref{fig:gmAdex_extDrug} and~\ref{fig:gmAdex_extFinInv}), has several bumps during the  
workday ($8$:$00$ - $16$:$00$), and is relatively low otherwise;   this again has implications for selecting ad content. 

\begin{figure}[htbp!]
\centering
\includegraphics[width=\figsize]{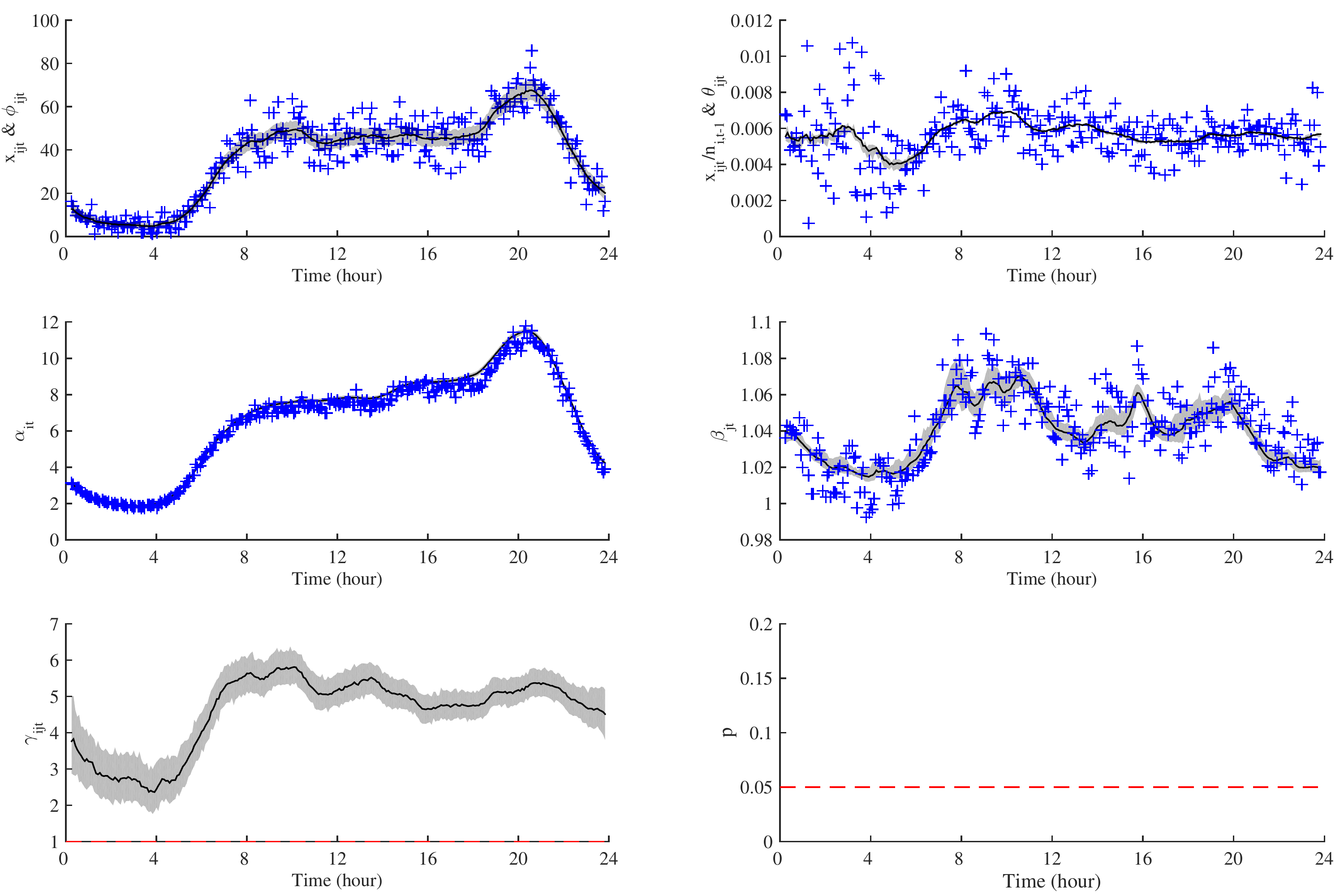} 
\caption{Posterior summaries for DGM parameters for transitions entering node $i=$ 
Shopping with details as in Fig.~\ref{fig:gmAdex_extWth}.}    
\label{fig:gmAdex_extShop} 
\bigskip\bigskip
\includegraphics[width=\figsize]{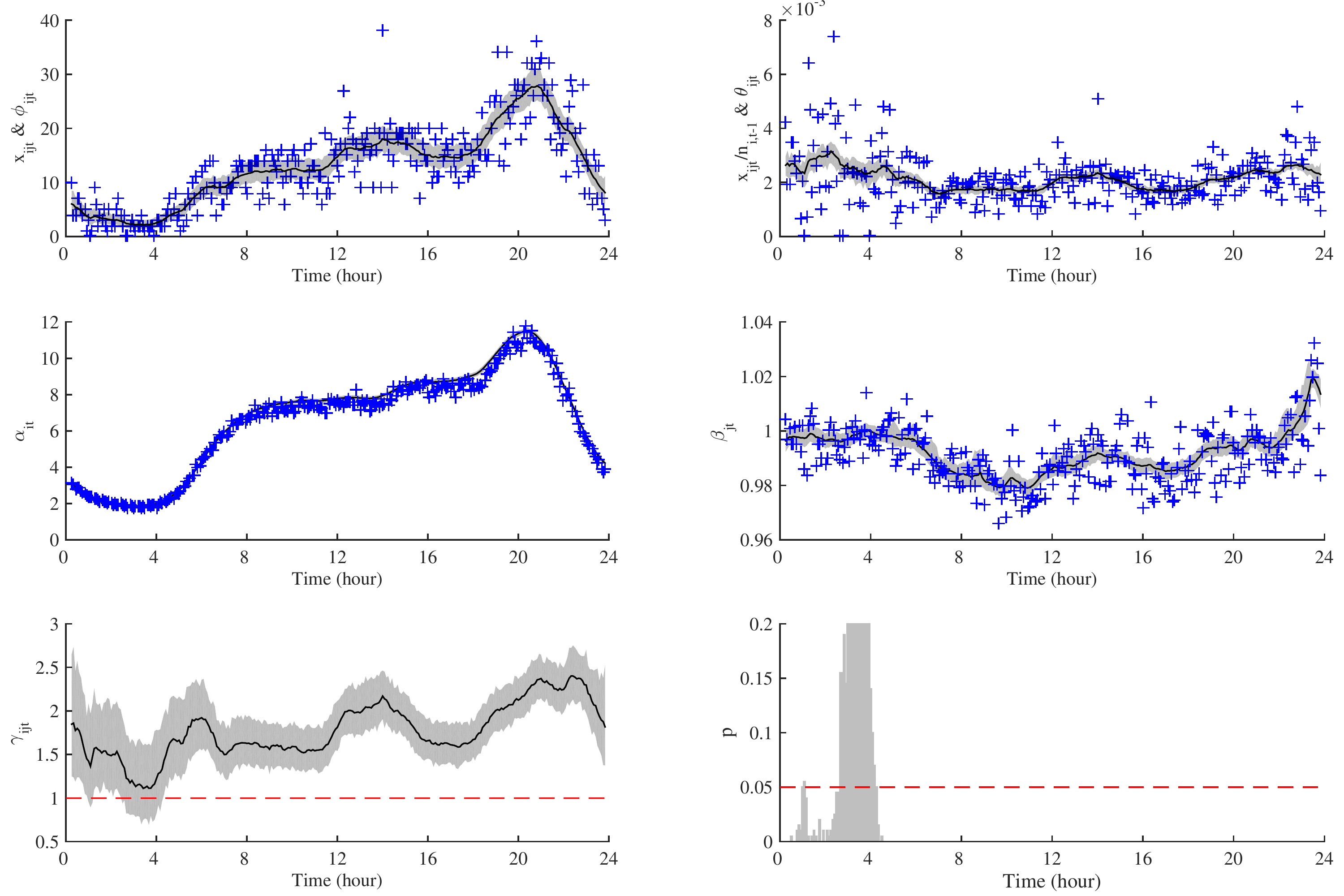} 
\caption{Posterior summaries for DGM parameters for transitions entering node $i=$ 
Beauty \& Fitness/Fashion \& Style with details as in 
Fig.~\ref{fig:gmAdex_extWth}.}
\label{fig:gmAdex_extFas}
\end{figure}

A further pattern in entering affinities has high intensity in the morning and then is stable during 
the day. 
The entering affinity of Shopping is a good example---see 
Fig.~\ref{fig:gmAdex_extShop}.
But things may be complex---perhaps someone on this website at a 
non-standard time is more apt to purchase than a casual browser.

One additional pattern in entering affinities is notable for multiple pronounced peaks. 
Beauty \& Fitness/Fashion \& Style (Fig.~\ref{fig:gmAdex_extFas})
has three peaks, at about $7$:$00$ - $8$:$00$, at about 
$12$:$00$ - $16$:$00$, and at about $20$:$00$ - $24$:$00$. 
September 17 is during the New York Fashion Week, which may lead to
atypical behavior.

\subsubsection*{Exiting affinities}
\begin{figure}[htbp!]
\centering
 \includegraphics[width=\figsize]{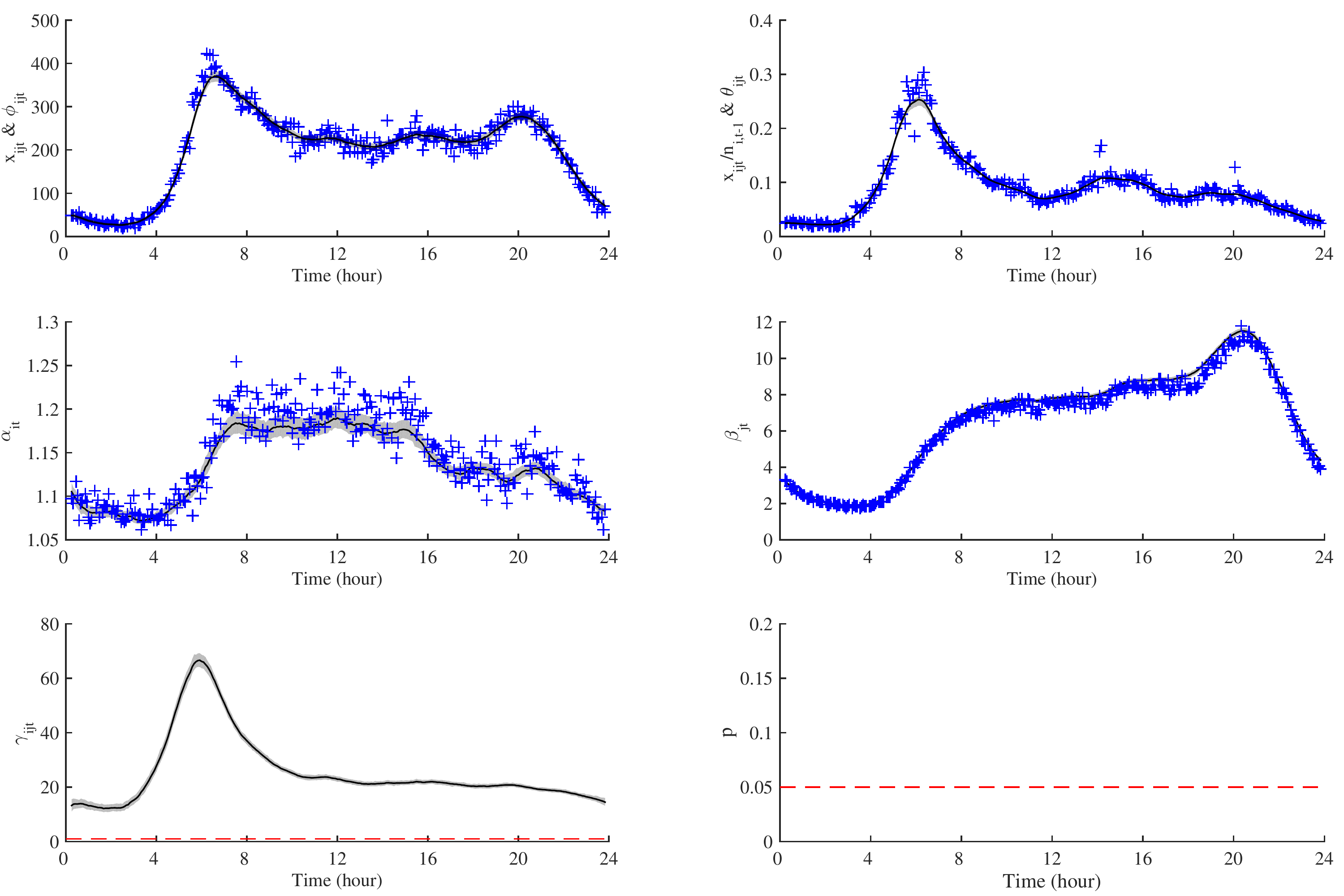} 
\caption{Posterior summaries for DGM parameters for transitions exiting from node $i=$ News/Weather with details as in Fig.~\ref{fig:gmAdex_Artext}.}    
\label{fig:gmAdex_Wthext}
\bigskip\bigskip

\includegraphics[width=\figsize]{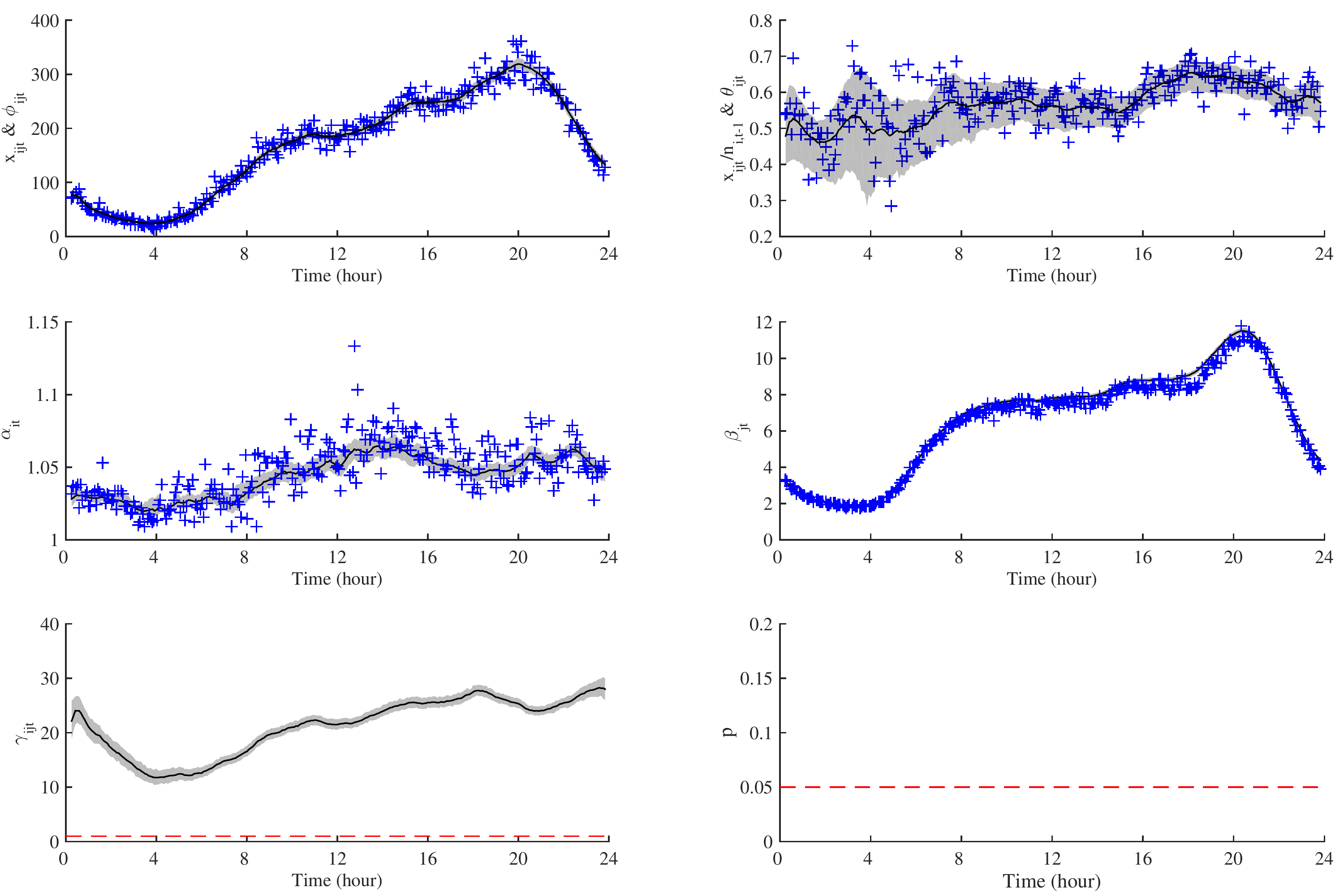} 
\caption{Posterior summaries for DGM parameters for transitions exiting from node $i=$ 
Arts \& Entertainment/Music \& Audio with details as in 
Fig.~\ref{fig:gmAdex_Artext}. Note the peak at night.}    
\label{fig:gmAdex_Musicext} 
\end{figure}

Other patterns have been detected in trajectories of exiting affinities. One has a peak, but is otherwise low, as 
in the exiting affinity for News/Weather (Fig.~\ref{fig:gmAdex_Wthext}). 
Clearly, people are visiting this site only to learn about the weather 
forecast, and then leave for work.

A third  pattern in trajectories of exiting affinities increases from $4$:$00$ in the morning until the evening, 
indicating that these categories are increasingly losing users throughout the day. 
Such categories include Arts \& Entertainment/Music \& Audio
(Fig.~\ref{fig:gmAdex_Musicext}).

\subsubsection*{Distinct node pair affinities}
\begin{figure}[htbp!]
\centering
\includegraphics[width=\figsize]{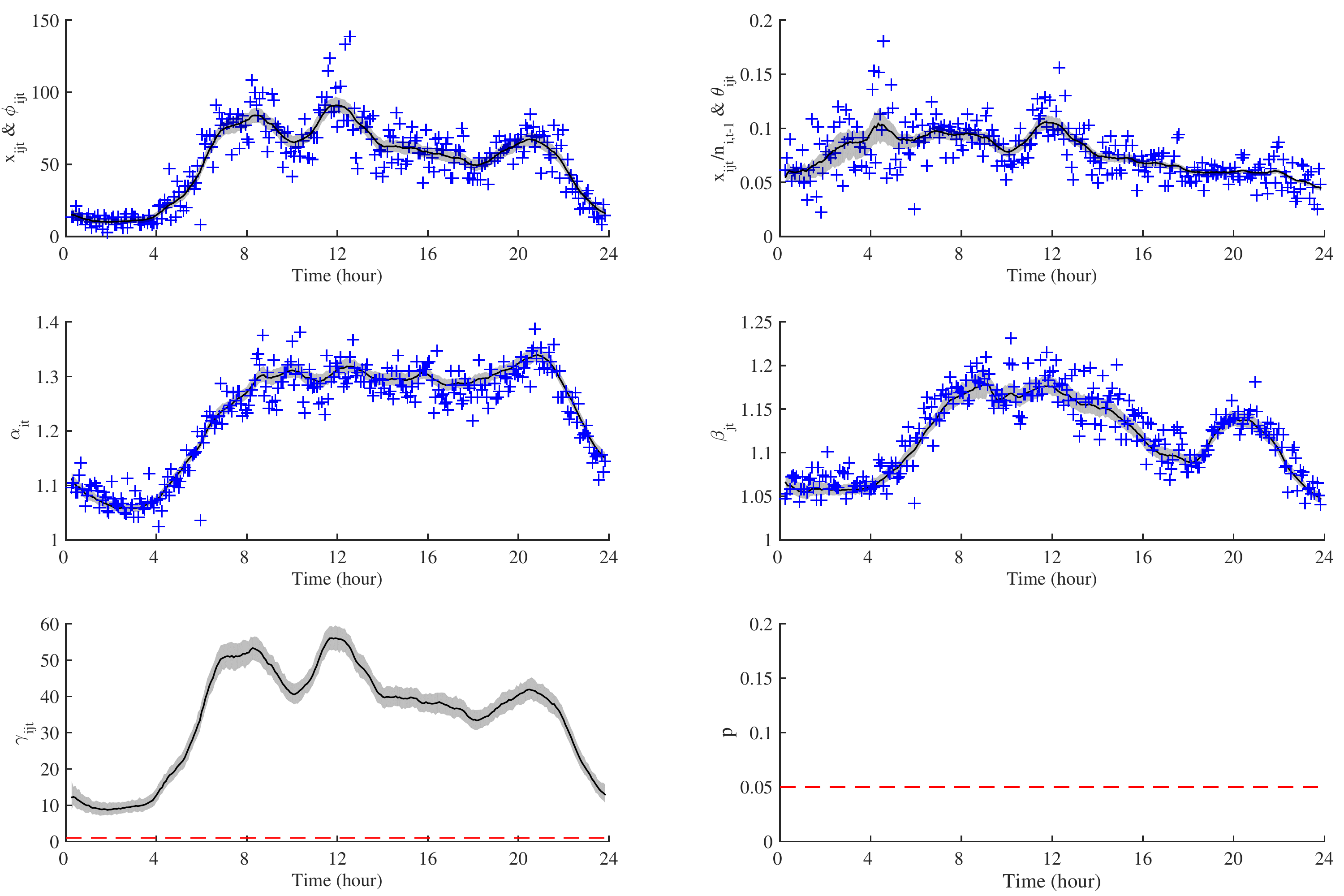} 
\caption{Posterior summaries for DGM parameters for transitions from node $i=$
News $\rightarrow j=$ News/Local News with details as in 
Fig.~\ref{fig:gmAdex_GamesVideo}.}    
\label{fig:gmAdex_NewsLocal}
\bigskip\bigskip
\includegraphics[width=\figsize]{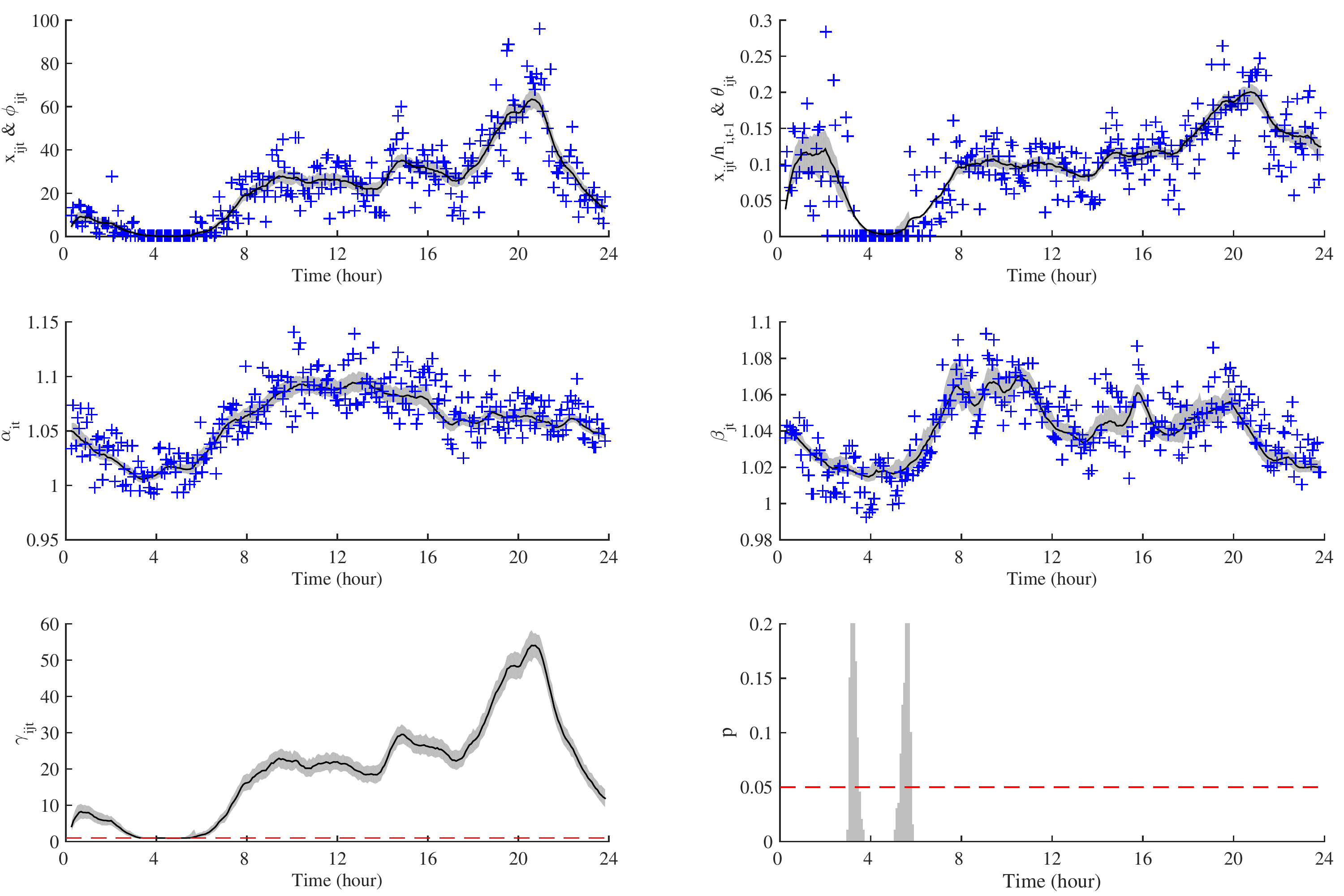} 
\caption{Posterior summaries for DGM parameters for transitions from node $i=$
News/Technology News $\rightarrow j=$ Shopping with details as in 
Fig.~\ref{fig:gmAdex_GamesVideo}.}    
\label{fig:gmAdex_TechShop} 
\end{figure}

The affinity from News to News/Local News
(Fig.~\ref{fig:gmAdex_NewsLocal}) has three peaks in the day. 
The first is around $8$:$00$, the second around noon, and last 
around $20$:$00$. 
The peaks indicate that people who are interested in national news 
also check the local news, and that the timing coincides with 
leisure.

The last example is the affinity from News/Technology News to 
Shopping (Fig.~\ref{fig:gmAdex_TechShop}). 
This peaks around $20$:$00$, which suggests the users who have read 
technology news start to explore technology purchase, and should be a clear signal for ad display.

\clearpage

\clearpage 
 
\section*{References}
\bibliographystyle{nws}
\renewcommand{\bibname}{}
\bibliography{ChenEtAl-DGLM-Networks}

\begin{thebibliography}{}

\bibitem[\protect\citename{Anacleto {\em et~al.}\relax,
  }2013a]{AnacletoEtAl2013a}
Anacleto, O., Queen, C., \& Albers, C.~J. (2013a).
\newblock Forecasting multivariate road traffic flows using {B}ayesian dynamic
  graphical models, splines and others traffic variables.
\newblock {\em {Australian and New Zealand Journal of Statistics}}, {\bf 55},
  69--86.

\bibitem[\protect\citename{Anacleto {\em et~al.}\relax,
  }2013b]{AnacletoEtAl2013b}
Anacleto, O., Queen, C., \& Albers, C.~J. (2013b).
\newblock Multivariate forecasting of road traffic flows in the presence of
  heteroscedasticity and measurement errors.
\newblock {\em {Journal of the Royal Statistical Society {\em (Series C:
  Applied Statistics)}}}, {\bf 62}, 251--270.

\bibitem[\protect\citename{Berry \& West, }2019]{BerryWest2018DCMM}
Berry, L.~R., \& West, M. (2019).
\newblock Bayesian forecasting of many count-valued time series.
\newblock {\em Journal of {B}usiness and {E}conomic {S}tatistics, to appear}.
\newblock arXiv:1805.05232.

\bibitem[\protect\citename{Berry {\em et~al.}\relax, }2018]{BerryWest2018TSM}
Berry, L.~R., Helman, P., \& West, M. (2018).
\newblock Probabilistic forecasting of heterogeneous consumer transaction-sales
  time series.
\newblock {\em International journal of forecasting, invited revise \&
  resubmit}.
\newblock arXiv:1808.04698.

\bibitem[\protect\citename{Bianchi {\em et~al.}\relax,
  }2018]{bianchi2018modeling}
Bianchi, D., Billio, M., Casarin, R., \& Guidolin, M. (2018).
\newblock Modeling systemic risk with {M}arkov switching graphical {SUR}
  models.
\newblock {\em {Journal of Econometrics}}.

\bibitem[\protect\citename{Chen {\em et~al.}\relax, }2018]{xi2017bdfm}
Chen, X., Irie, K., Banks, D., Haslinger, R., Thomas, J., \& West, M. (2018).
\newblock Scalable {B}ayesian modeling, monitoring and analysis of dynamic
  network flow data.
\newblock {\em Journal of the {A}merican {S}tatistical {A}ssociation}, {\bf
  113}, 519--533.

\bibitem[\protect\citename{Congdon, }2000]{Cogdon2000}
Congdon, P. (2000).
\newblock A {B}ayesian approach to prediction using the gravity model, with an
  application to patient flow modeling.
\newblock {\em Geographical {A}nalysis}, {\bf 32}, 205--224.

\bibitem[\protect\citename{Giot {\em et~al.}\relax, }2003]{giot2003protein}
Giot, L., Bader, J.~S., Brouwer, C., Chaudhuri, A., Kuang, B., Li, Y., Hao,
  Y.~L., Ooi, C.~E., Godwin, B., \& Vitols, E.~{\em et al}. (2003).
\newblock A protein interaction map of {D}rosophila {M}elanogaster.
\newblock {\em Science}, {\bf 302}, 1727--1736.

\bibitem[\protect\citename{Giraitis {\em et~al.}\relax,
  }2016]{giraitis2016estimating}
Giraitis, L., Kapetanios, G., Wetherilt, A., \& {\v{Z}}ike{\v{s}}, F. (2016).
\newblock Estimating the dynamics and persistence of financial networks, with
  an application to the {S}terling money market.
\newblock {\em {Journal of Applied Econometrics}}, {\bf 31}, 58--84.

\bibitem[\protect\citename{Goldstein, }1976]{goldstein1976bayesian}
Goldstein, M. (1976).
\newblock Bayesian analysis of regression problems.
\newblock {\em Biometrika}, {\bf 63}, 51--58.

\bibitem[\protect\citename{Gruber \& West, }2016]{GruberWest2016BA}
Gruber, L.~F., \& West, M. (2016).
\newblock {GPU}-accelerated {B}ayesian learning in simultaneous graphical
  dynamic linear models.
\newblock {\em Bayesian {A}nalysis}, {\bf 11}, 125--149.

\bibitem[\protect\citename{Gruber \& West, }2017]{GruberWest2017ECOSTA}
Gruber, L.~F., \& West, M. (2017).
\newblock Bayesian forecasting and scalable multivariate volatility analysis
  using simultaneous graphical dynamic linear models.
\newblock {\em Econometrics and {S}tatistics}, {\bf 3}, 3--22.

\bibitem[\protect\citename{Hanneke {\em et~al.}\relax,
  }2010]{hanneke2010discrete}
Hanneke, S., Fu, W., \& Xing, E.~P. (2010).
\newblock Discrete temporal models of social networks.
\newblock {\em {Electronic Journal of Statistics}}, {\bf 4}, 585--605.

\bibitem[\protect\citename{Hartigan, }1969]{hartigan1969linear}
Hartigan, J.~A. (1969).
\newblock Linear {B}ayesian methods.
\newblock {\em Journal of the {R}oyal {S}tatistical {S}ociety ({S}eries {B}:
  {M}ethodological)}, {\bf 31}, 446--454.

\bibitem[\protect\citename{Hoff, }2011]{hoff2011hierarchical}
Hoff, P.~D. (2011).
\newblock Hierarchical multilinear models for multiway data.
\newblock {\em {Computational Statistics and Data Analysis}}, {\bf 55},
  530--543.

\bibitem[\protect\citename{Holme, }2015]{holme2015modern}
Holme, P. (2015).
\newblock Modern temporal network theory: A colloquium.
\newblock {\em {European Physical Journal B}}, {\bf 88}, 234.

\bibitem[\protect\citename{Holme \& Saram{\"a}ki, }2013]{holme2013temporal}
Holme, P., \& Saram{\"a}ki, J. (2013).
\newblock {\em Temporal {N}etworks}.
\newblock Springer.

\bibitem[\protect\citename{Jansen {\em et~al.}\relax, }2007]{Jansen2007}
Jansen, B.~J., Spink, A., \& Kathuria, V. (2007).
\newblock How to define searching sessions on web search engines.
\newblock {\em Pages  92--109 of:} Nasraoui, O., Spiliopoulou, M., Srivastava,
  J., Mobasher, B., \& Masand, B. (eds), {\em {Advances in Web Mining and Web
  Usage Analysis: Eighth International Workshop on Knowledge Discovery on the
  Web, WebKDD 2006}}.
\newblock Lecture Notes in Computer Science.
\newblock Springer.

\bibitem[\protect\citename{Kim {\em et~al.}\relax, }2018]{kim2018review}
Kim, B., Lee, K.~H., Xue, L., \& Niu, X. (2018).
\newblock A review of dynamic network models with latent variables.
\newblock {\em Statistics {S}urveys}, {\bf 12}, 105--135.

\bibitem[\protect\citename{Koren {\em et~al.}\relax, }2009]{Koren:2009}
Koren, R., Bell, R., \& Volinsky, C. (2009).
\newblock Matrix factorization techniques for recommender systems.
\newblock {\em Computer}, {\bf 8}, 30--37.

\bibitem[\protect\citename{McCullough \& Nelder,
  }1989]{mccullough1989generalized}
McCullough, P., \& Nelder, J.~A. (1989).
\newblock {\em Generalized {L}inear {M}odels}.
\newblock Chapman \& Hall.

\bibitem[\protect\citename{Migon \& Harrison, }1985]{migon1985application}
Migon, H.~S., \& Harrison, P.~J. (1985).
\newblock An application of non-linear {B}ayesian forecasting to television
  advertising.
\newblock {\em Pages  681--696 of:} Bernardo, J.~M., DeGroot, M.~H., Lindley,
  D.~V., \& Smith, A. F.~M. (eds), {\em Bayesian {S}tatistics 2}.
\newblock North-Holland, Amsterdam, and Valencia University Press.

\bibitem[\protect\citename{Newman, }2004]{newman2004analysis}
Newman, M. E.~J. (2004).
\newblock Analysis of weighted networks.
\newblock {\em {Physical Review E}}, {\bf 70}, 056131.

\bibitem[\protect\citename{Newman, }2018]{newman2018network}
Newman, M. E.~J. (2018).
\newblock Network structure from rich but noisy data.
\newblock {\em Nature {P}hysics}, {\bf 14}, 542.

\bibitem[\protect\citename{Pang \& Lee, }2008]{Pang:2008}
Pang, B., \& Lee, L. (2008).
\newblock Opinion mining and sentiment analysis.
\newblock {\em {Foundations and Trends in Information Retrieval}}, {\bf 2},
  1--135.

\bibitem[\protect\citename{Prado \& West, }2010]{PradoWest2010}
Prado, R., \& West, M. (2010).
\newblock {\em {Time Series: Modeling, Computation and Inference}}.
\newblock Chapman \& Hall/CRC Press.

\bibitem[\protect\citename{Queen \& Albers, }2009]{Queen2009}
Queen, C.~M., \& Albers, C.~J. (2009).
\newblock Intervention and causality: {F}orecasting traffic flows using a
  dynamic {B}ayesian network.
\newblock {\em {Journal of the American Statistical Association}}, {\bf 104},
  669--681.

\bibitem[\protect\citename{Richard {\em et~al.}\relax, }2014]{richard2014link}
Richard, E., Ga{\"\i}ffas, S., \& Vayatis, N. (2014).
\newblock Link prediction in graphs with autoregressive features.
\newblock {\em {Journal of Machine Learning Research}}, {\bf 15}, 565--593.

\bibitem[\protect\citename{Sarkar {\em et~al.}\relax, }2007]{sarkar2007latent}
Sarkar, P., Siddiqi, S.~M., \& Gordon, G.~J. (2007).
\newblock A latent space approach to dynamic embedding of co-occurrence data.
\newblock {\em Pages  420--427 of:} {\em {Artificial Intelligence and
  Statistics}}.

\bibitem[\protect\citename{Sen \& Smith, }1995]{Sen:1995}
Sen, A., \& Smith, T. (1995).
\newblock {\em {Gravity Models of Spatial Interaction Behavior}}.
\newblock Springer.

\bibitem[\protect\citename{Soriano {\em et~al.}\relax, }2013]{Soriano:2013}
Soriano, J., Au, T., \& Banks, D. (2013).
\newblock Text mining in computational advertising.
\newblock {\em {Statistical Analysis and Data Mining}}, {\bf 6}, 273--285.

\bibitem[\protect\citename{Tebaldi \& West, }1998]{Tebaldi1998}
Tebaldi, C., \& West, M. (1998).
\newblock Bayesian inference on network traffic using link count data.
\newblock {\em Journal of the {A}merican {S}tatistical {A}ssociation}, {\bf
  93}, 557--573.

\bibitem[\protect\citename{Tebaldi {\em et~al.}\relax, }2002]{Tebaldi2002}
Tebaldi, C., West, M., \& Karr, A.~F. (2002).
\newblock Statistical analyses of freeway traffic flows.
\newblock {\em Journal of {F}orecasting}, {\bf 21}, 39--68.

\bibitem[\protect\citename{Uetz {\em et~al.}\relax,
  }2000]{uetz2000comprehensive}
Uetz, P., Giot, L., Cagney, G., Mansfield, T.~A., Judson, R.~S., Knight, J.~R.,
  Lockshon, D., Narayan, V., Srinivasan, M., \& Pochart, P.~{\em et al}.
  (2000).
\newblock A comprehensive analysis of protein-protein interactions in
  saccharomyces cerevisiae.
\newblock {\em Nature}, {\bf 403}, 623.

\bibitem[\protect\citename{West, }1985]{west1985generalized}
West, M. (1985).
\newblock Generalized linear models: scale parameters, outlier accommodation
  and prior distributions (with discussion).
\newblock {\em Pages  531--558 of:} Bernardo, J.~M., DeGroot, M.~H., Lindley,
  D.~V., \& Smith, A. F.~M. (eds), {\em Bayesian {S}tatistics 2}.
\newblock North-Holland, Amsterdam, and Valencia University Press.

\bibitem[\protect\citename{West, }1994]{West1994}
West, M. (1994).
\newblock {\em Statistical inference for gravity models in transportation flow
  forecasting}.
\newblock Discussion Paper 94-20, Duke University, and Technical Report \#60,
  National Institute of Statistical Sciences.

\bibitem[\protect\citename{West \& Harrison, }1997]{WestHarrison1997}
West, M., \& Harrison, P.~J. (1997).
\newblock {\em {Bayesian Forecasting and Dynamic Models}}. 2nd edn.
\newblock Springer.

\bibitem[\protect\citename{West {\em et~al.}\relax, }1985]{west1985dynamic}
West, M., Harrison, P.~J., \& Migon, H.~S. (1985).
\newblock Dynamic generalized linear models and {B}ayesian forecasting (with
  discussion).
\newblock {\em {Journal of the American Statistical Association}}, {\bf 80},
  73--83.

\bibitem[\protect\citename{Xing {\em et~al.}\relax, }2010]{xing2010state}
Xing, E.~P., Fu, W., \& Song, L. (2010).
\newblock A state-space mixed membership block model for dynamic network
  tomography.
\newblock {\em {Annals of Applied Statistics}}, {\bf 4}, 535--566.

\bibitem[\protect\citename{Xu \& Hero, }2014]{xu2014dynamic}
Xu, K.~S., \& Hero, A.~O. (2014).
\newblock Dynamic stochastic block models for time-evolving social networks.
\newblock {\em {IEEE Journal of Selected Topics in Signal Processing}}, {\bf
  8}, 552--562.

\end{thebibliography}

\end{document}